\documentclass[aps, prl, twocolumn, nofootinbib, floatfix,superscriptaddress]{revtex4-2}

\usepackage{graphicx}
\usepackage{dcolumn}
\usepackage{color}
\usepackage{hyperref}   % add hypertext capabilities

\usepackage[dvips]{epsfig}
\usepackage{amssymb}

\usepackage{amsmath}
\usepackage{xcolor}

\newcommand{\rv}[1]{#1}

\usepackage{eucal}
\usepackage{mathrsfs}
\usepackage{amsmath}
\usepackage{bm}
\usepackage{amssymb}
\usepackage{amsthm}
\usepackage{amsfonts}
\usepackage{latexsym}
\usepackage{mathtools}

\usepackage{chemformula}    % chemistry notation (APS wants Roman Upright font)
\usepackage{siunitx}
\DeclareSIUnit\Molar{\textsc{m}}

\theoremstyle{definition}

\definecolor{MidnightBlue}{HTML}{191970}
\hypersetup{
	colorlinks=true,
	allcolors=MidnightBlue
}

\usepackage{verbatim}

\begin{document}
%TC:ignore

\title{Discovering dynamics and parameters of nonlinear oscillatory and chaotic systems from partial observations}

\author{George Stepaniants}
\affiliation{Department of Mathematics, Massachusetts Institute of Technology, Cambridge, Massachusetts 02139-4307, USA}
\author{Alasdair D. Hastewell}
\affiliation{Department of Mathematics, Massachusetts Institute of Technology, Cambridge, Massachusetts 02139-4307, USA}
\author{Dominic J. Skinner}
\affiliation{Department of Mathematics, Massachusetts Institute of Technology, Cambridge, Massachusetts 02139-4307, USA}
\affiliation{NSF-Simons Center for Quantitative Biology, Northwestern University, Evanston, Illinois 60208, USA}
\author{Jan F. Totz}
\affiliation{Department of Mathematics, Massachusetts Institute of Technology, Cambridge, Massachusetts 02139-4307, USA}
\author{J\"orn Dunkel}
\affiliation{Department of Mathematics, Massachusetts Institute of Technology, Cambridge, Massachusetts 02139-4307, USA}

\date{\today}

\begin{abstract}
 Despite rapid  progress in live-imaging techniques, many complex biophysical and biochemical systems remain only  partially observable, thus posing the challenge to identify valid theoretical models and estimate their parameters  from an incomplete set of experimentally accessible time series.
 Here, we combine sensitivity methods and \rv{ranked-choice model selection} to construct an automated hidden dynamics inference framework that can discover predictive nonlinear dynamical models for both observable and latent variables from noise-corrupted incomplete data in oscillatory and chaotic systems.  After validating the framework for prototypical FitzHugh-Nagumo oscillations, we demonstrate its applicability to experimental data from  squid neuron activity measurements and Belousov-Zhabotinsky (BZ)  reactions, as well as to the Lorenz system in the chaotic regime. 
 \end{abstract}

\maketitle

%TC:endignore

%%
%
%
Nonlinear oscillations and chaos are ubiquitous in natural and man-made systems~\cite{strogatz2018nonlinear}, % found in practically every real-world system, 
from neurons~\cite{stiefel2016neurons, buzsaki2004neuronal} and biochemical networks~\cite{Srinivas2017} to power grids~\cite{filatrella2008analysis, rohden2012self}, lasers~\cite{Marty2021}, and the  Earth's climate~\cite{Vettoretti2022}. Major advances in live-imaging and fluorescence labeling techniques  over the last decades have made it possible to record extensive time-series data of neuronal~\cite{ling2020high, atanas2022neural} and other cellular activity~\cite{jeckel2019learning, alvelid2022event} at high temporal resolution. Yet, notwithstanding such progress, for many complex biophysical and biochemical systems, direct measurements are limited to a single experimentally accessible observable~\cite{xie2015intracellular} while essential components of the underlying dynamical circuit stay hidden~\cite{Paulsson2016}. Limited observability has led to the emergence of competing theoretical models for neuronal~\cite{IZH07} and gene-regulatory networks~\cite{Paulsson2016}, and identifying valid models and their parameters from incomplete data remains a central challenge. Here, we combine sensitivity methods~\cite{rackauckas2020universal} for differential equations  with ranked choice voting~\cite{kemeny1959mathematics, young1995optimal} to construct a hidden dynamics inference (HDI) framework that can discover predictive nonlinear dynamical models for both observable and latent variables from noise-corrupted incomplete data in oscillatory and chaotic systems.
\par
Driven by the rapidly advancing data acquisition techniques, dynamical model inference is becoming increasingly more important\rv{~\cite{aguirre2009modeling, brunton2016discovering}} in climate physics~\cite{talagrand1987variational, courtier1987variational, majda2009normal}, fluid mechanics~\cite{raissi2018hidden, Brenner2021} and biophysics~\cite{yang2021inference, supekar2021learning, romeo2021learning}. Time-delay embeddings~\cite{takens1981detecting, sugihara2012detecting}, recurrent neural networks~\cite{haehne2019detecting} and autoencoders~\cite{chen2021discovering} have successfully been used to estimate hidden dimensions and forecast complex dynamics~\cite{hewamalage2021recurrent}, but such \lq equation-free\rq{} approaches often cannot reveal coupling mechanisms and their dependencies on experimental conditions. Complementary equation-based approaches~\cite{brunton2016discovering} have shown promise in learning interpretable dynamical models from partially observed data using physics-informed neural networks~\cite{bakarji2022discovering, lu2022discovering, raissi2020hidden, ouala2020learning, ayed2019learning}, manifold methods~\cite{cenedese2022data}, or data assimilation~\cite{ribera2022model}, enabling prediction of nonlinear and chaotic dynamics in mechanical, electrical, and hydrodynamic systems \rv{(see SI for comprehensive discussion)}. Despite such substantial progress, however, applications to experimental data from nonlinear biophysical and biochemical systems still face many open problems, as existing methods require long time series recordings with low noise (e.g. to construct time-delay embeddings or train neural networks) and do not ensure stability of learned models.
\par
The HDI framework introduced here overcomes these challenges by integrating the robustness of sensitivity methods~\cite{rackauckas2020universal} and ranked-choice model selection~\cite{kemeny1959mathematics, young1995optimal} with traditional library-based learning methods~\cite{brunton2016discovering, fasel2022ensemble}. This enables us to learn physically interpretable models for partially hidden nonlinear systems from short, highly noisy data trajectories in a manner that ensures correct long time dynamics. Since the hidden-variable dynamical equation discovered from partial observations may not be unique, we develop a systematic algebraic procedure (SI) for comparing learned models. After validating the HDI framework on strongly noise corrupted simulations of the FitzHugh-Nagumo oscillator, we apply our approach to experimental measurements of squid neuron spike trains and video observations of Belousov-Zhabotinsky chemical reactions, demonstrating how HDI can be used to measure model parameters as a function of external experimental conditions.

\begin{figure*}
    \centering
    \includegraphics{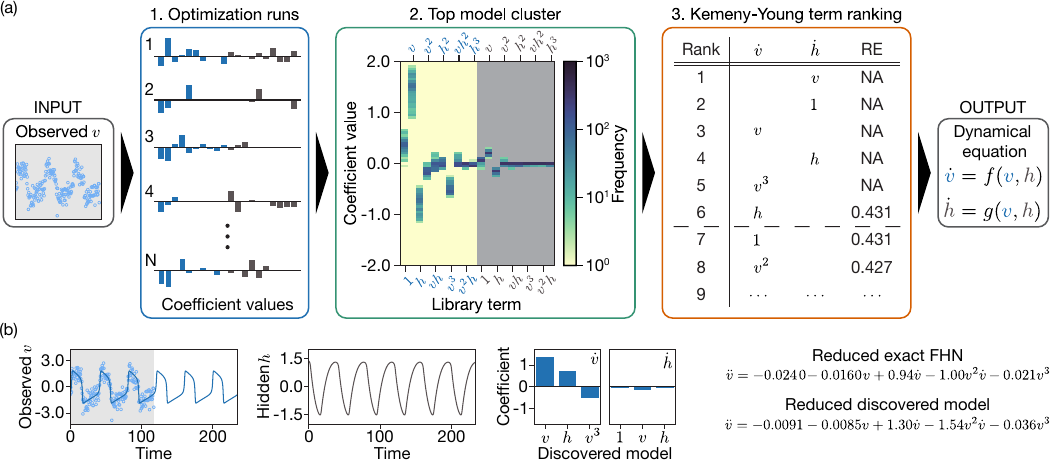}
    \caption{
    General HDI framework illustrated for strongly noise-corrupted FHN simulation data.
    (a)~Algorithm flow chart:  
    (1)~ODE sensitivity optimization~\cite{rackauckas2020universal} yields $N\sim 20,000$ candidate models by \rv{tuning $20$ parameters of dense two field cubic observed (blue) and hidden (dark-gray) variable equations from random initializations (SI)}. Models are filtered for stability and fit quality~(SI).
    (2) The remaining $\sim 4000$ models are hierarchically clustered using the cosine similarity between their parameter vectors~(SI). Histograms of parameters in the largest cluster are used to rank the terms based on their coefficient of variation~(SI).
    (3) \rv{Kemeny-Young} ranking produces a list of candidate models of decreasing sparsity. Models are refit at each sparsity level, and the user can select the model that best balances sparsity and relative error (RE).
    (b) Using data from only the $v$ time series corrupted by 50\% noise, HDI correctly discovers a sparse first-order system that reduces to the same second-order form as the FHN model.}
    \label{fig:fig1_fhn}
\end{figure*}

\begin{figure*}
    \centering
    \includegraphics{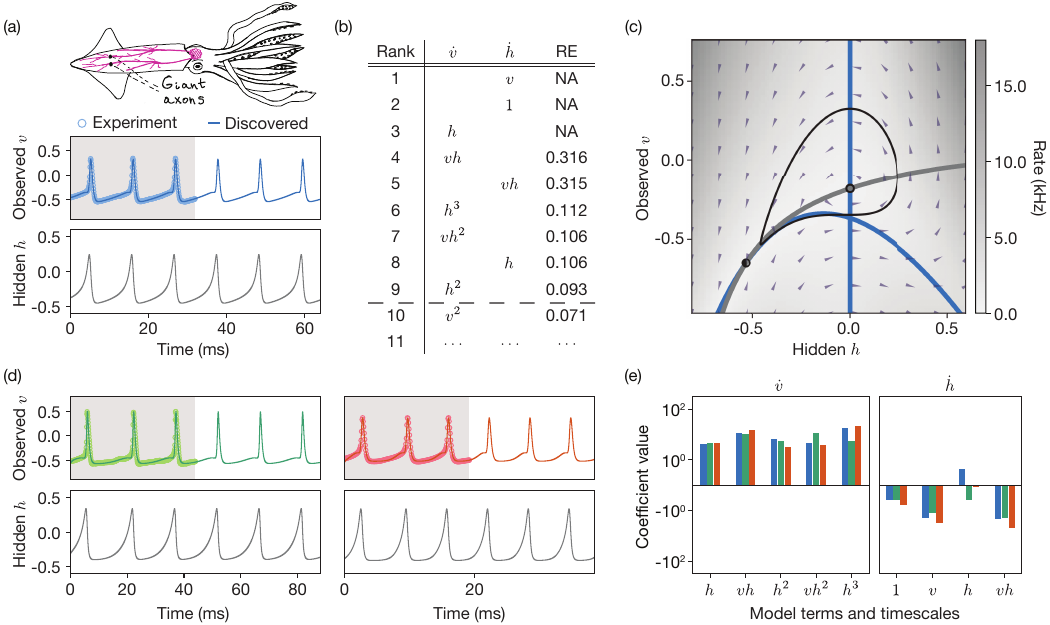}
    \caption{
    HDI framework learns a parsimonious two-variable model from an experimental recording of the membrane potential in a squid giant axon and reproduces the dynamics in additional squid giant axons from the SGAMP database~\cite{paydarfar2006noisy, physionet2000}.
    (a)~North Atlantic longfin inshore squid (\textit{Loligo pealeii}) with sketch of the nervous system and position of giant axons (top). Learned two-variable HDI model with 9 terms accurately fits the membrane potential $v$ (center, line) of an experimental squid giant axon (open circles) in response to a noisy stimulus input current. The hidden variable $h$ (bottom) acts as a slow recovery variable.
    (b)~Polynomial model terms in $\dot{v}$ and $\dot{h}$ equations ranked from most to least important based on their coefficient of variation in the largest model cluster. \rv{Training} data losses of sparse models containing only top $s$ ranked terms are shown and model with sparsity nine is chosen.
    (c) Limit cycle and fixed points (black) of learned model are consistent with prior models of regular spiking neurons~\cite{IZH07} where the \rv{proximity of the saddle fixed point} to the orbit likely arises from a homoclinic bifurcation. Nullclines of $v, h$ plotted in blue and gray respectively.
    (d) Selected nine term model (line) generalizes to two additional squid axon recordings (open circles).
    (e) Coefficients of the nine term model align across all three train and test squid axon experiments.}
    \label{fig:fig2_neuron}
\end{figure*}

\begin{figure*}
    \centering
    \includegraphics{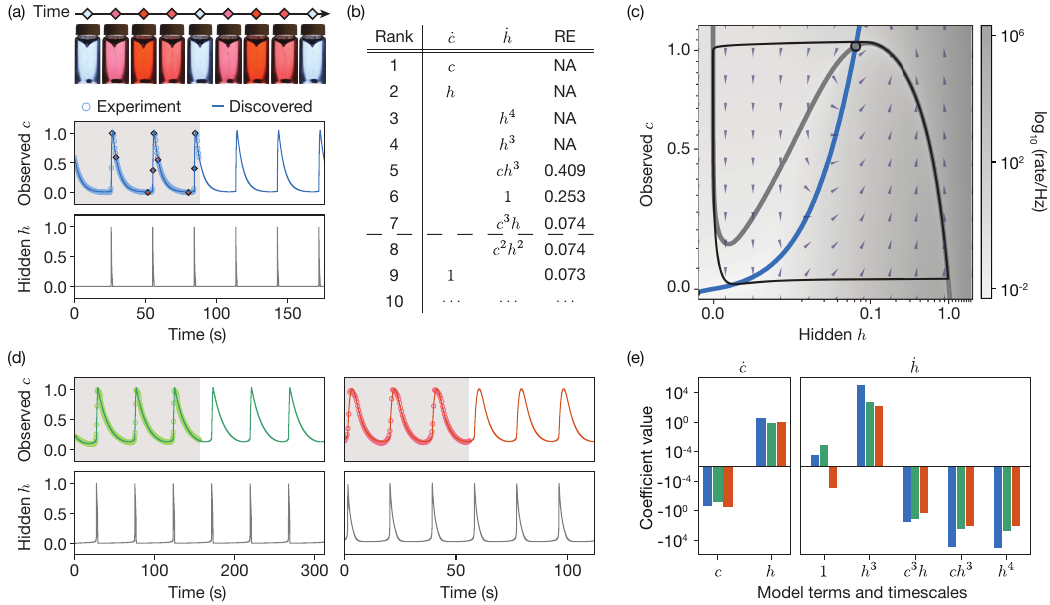}
    \caption{HDI applied to our experimental BZ reaction data learns a two-variable linear-quartic model that generalizes under catalyst variations. 
    (a)~Experimental snapshots of the BZ reaction showing periodic  color oscillations (top). Input data (open circles) and observed and hidden variables (solid line) integrated from the learned polynomial ODE model. Using data from three oscillations the learning framework finds that a seven term ODE can accurately describe the dynamics.
    (b) Polynomial ODE terms appearing in $\dot{c}$ and $\dot{h}$ equations ranked from most to least important based on their coefficient of variation. Model terms are added one-at-a-time in order of importance with the seventh term leading to a drop in the training loss.
    (c) Phase plane diagram of learned seven term ODE from previous panel contains crucial features found in most two-variable BZ models~\cite{TYS82}. Limit cycle contains an unstable fixed point (black) with \rv{a monotonic $x$-nullcline (blue) and an} $h$-nullcline (dark gray) in the form of a ``cubic" curve as found in the FHN, Rovinsky and ZBKE models.
    (d) Resulting seven-term model (solid line) accurately fits the dynamics of the color of the chemical solution (open circles) in two new BZ experiments.
    (e) Coefficients of the model remain consistent across all three experimental BZ reactions.
    Chemical concentrations:
    \SI{0.20}{\Molar}\,\ch{H2SO4}, \SI{0.11}{\Molar}\,\ch{NaBrO3}, \SI{0.05}{\Molar}\,\ch{CH2(COOH)2}, \SI{0.03}{\Molar}\,\ch{NaBr}, \SI{0.3}{\milli\Molar}\,ferroin (blue),
    \SI{0.41}{\Molar}\,\ch{H2SO4}, \SI{0.17}{\Molar}\,\ch{NaBrO3}, \SI{0.03}{\Molar}\,\ch{CH2(COOH)2}, \SI{0.02}{\Molar}\,\ch{NaBr}, \SI{0.3}{\milli\Molar}\,ferroin (green),
    \SI{0.51}{\Molar}\,\ch{H2SO4}, \SI{0.10}{\Molar}\,\ch{NaBrO3}, \SI{0.03}{\Molar}\,\ch{CH2(COOH)2}, \SI{0.02}{\Molar}\,\ch{NaBr}, \SI{0.3}{\milli\Molar}\,ferroin (red).
    }
    \label{fig:fig3_bz}
\end{figure*}

A canonical example of a nonlinear oscillator is the FitzHugh-Nagumo (FHN) model \cite{IZH07}
\begin{equation}\label{eq:fhn}
    \dot{v} = v - \frac{v^3}{3} - w + I, \quad \dot{w} = \tau\Big(v + a - bw\Big),
\end{equation}
a simplified model of a firing neuron where the membrane voltage $v$ undergoes a rapid increase before  being diminished by the slow recovery variable $w$~\cite{fitzhugh1961impulses}. The rapid spiking and slow recovery arises from a separation in time scales $\tau \ll 1$ between variables. FHN has become a prototypical model of neuron spike trains, as it is stable and parsimonious, \rv{relying only on a small} number of polynomial terms. The HDI framework aims to learn models of this type from limited noisy recordings of a single variable, for example the $v$-coordinate of FHN [Fig.~\ref{fig:fig1_fhn}(a)]. This motivates us to define the following class of models, with observed variables $x_1, \hdots, x_m$ and hidden variables $h_{m+1}, \dots, h_M$, given by
\begin{subequations} \label{eq:hfm}
\begin{align}
    \dot{x}_k &= \tau_k\sum_{|\bm\alpha| \leq d_k} c_{\bm\alpha}^k x_1^{\alpha_1}\hdots h_M^{\alpha_M}, \quad 1 \leq k \leq m\\
    \dot{h}_k &= \tau_k\sum_{|\bm\alpha| \leq d_k} c_{\bm\alpha}^k x_1^{\alpha_1} \hdots h_M^{\alpha_M}, \quad m < k \leq M
\end{align}
\end{subequations}
which encompass a broad range of nonlinear oscillatory dynamics. Here, we only use polynomial terms on the right-hand side of the equation, although this can be extended to any other nonlinearities\rv{, such as trigonometric functions (SI)}. To avoid scaling ambiguities between $\tau_k$ and $\mathbf{c}^k = \{c_{\bm\alpha}^k\}$ in Eq.~\eqref{eq:hfm} we enforce that each $\mathbf{c}^k$ has unit norm. HDI models are determined by a parameter vector $\mathbf{p}$ containing the initial conditions of the variables $\{x_k^0\}_{k=1}^m, \{h_k^0\}_{k=m+1}^M$, time scales $\{\tau_k\}_{k=1}^M$, and polynomial coefficients $\{\mathbf{c}^k\}_{k=1}^M$. \rv{While time-delay embeddings can be used to provide lower bound estimates on the number of hidden variables $M-m$, here we restrict to periodic models with $M = 2$ variables or chaotic models with $M = 3$ variables, which we find sufficient to explain the experimental data.}

\par
To demonstrate the HDI framework (Fig.~\ref{fig:fig1_fhn}),  we consider noise-corrupted observations $y_{i1} = v(t_i) + \xi_{i1}$ of the $v$-coordinate of the FHN model~\eqref{eq:fhn}  [Fig.~\ref{fig:fig1_fhn}(a, INPUT)]. HDI repeatedly fits hidden two-variable models $(x_1, h_1)$ by minimizing the mean square error on the observed variables
\begin{equation}\label{eq:obj}
    \text{MSE}(\mathbf{p}) = \frac{1}{n}\sum_{i=1}^n\sum_{k=1}^m(x_k(t_i, \mathbf{p}) - y_{ik})^2
\end{equation}
plus a regularization term $\text{Reg}(\mathbf{p})$ that enforces the unit norm constraint on $\mathbf{c}^k$ and promotes sparsity, favoring lower-order terms that often lead to more stable dynamics (SI).  The full objective function $L(\mathbf{p}) = \text{MSE}(\mathbf{p}) + \text{Reg}(\mathbf{p})$ is minimized using ODE sensitivities and gradient descent methods~\cite{zhuang2020adabelief, fletcher2013practical} from random initializations [Fig.~\ref{fig:fig1_fhn}(a, 1)]. Noise robustness in our approach comes from using the full ODE solution $x_k(t_i, \mathbf{p})$ in the objective function, which avoids numerically differentiating noisy time series data, a typically ill-posed problem~\cite{van2020numerical}, and enforces that models learned are stable over the time-span of the training data. Multiple fits are required to sufficiently sample multiple local minima of the complex non-convex loss landscape $L(\mathbf{p})$: the model described by $\mathbf{p}$ is not unique since model symmetries -- linear, $h \mapsto \alpha h + \beta$, and nonlinear transformations of the hidden variables -- produce new models with identical $x_k$ dynamics (SI).
\par
To select a single candidate model, a set of models are optimized from random initialization. Model quality is measured using the relative error $\text{RE}(\mathbf{p}) = \sqrt{\text{MSE}(\mathbf{p}) / \text{Var}(\{y_{i1}\}_{i=1}^n)}$, given here for a single ($m = 1$) observable, where $\text{Var}$ is the uncorrected sample variance. Outlier models with incorrect dynamics (nonperiodic, nonchaotic, divergent, etc.) or large RE values are removed in an automated manner, and the remaining models are hierarchically clustered using the cosine similarity between their parameter vectors taking into account possible linear hidden variable transformations (SI). Measurement noise and regularization break many of the symmetry ambiguities resulting in a dominant largest cluster; on FHN data corrupted by $50\%$ noise we start by optimizing $20,000$ dense two-variable cubic models resulting in 4,006 filtered models of which 427 models form the dominant cluster [Fig.~\ref{fig:fig1_fhn}(a, 2)]. Sparse models are identified from the dominant cluster by ranking each term by its coefficient of variation in the cluster, the interquartile range divided by the median. Rankings are aggregated over a range of clustering thresholds using \rv{the Kemeny-Young method} to provide a robust ordering of terms. Based on this ordering,  a list of candidate models of decreasing sparsity containing the top ranked $s$ terms can be refit [Fig.~\ref{fig:fig1_fhn}(a, 3)]. From this list, practitioners can determine a suitable model sparsity that balances the trade-off between a model's complexity and RE based on their scientific judgement [Fig.~\ref{fig:fig1_fhn}(a, OUTPUT)].
\par
From just three noisy oscillations of the FHN $v$-coordinate, we learn a list of two-variable HDI models that at sparsities six and seven recovers models which are equivalent to FHN [Fig.~\ref{fig:fig1_fhn}(a, 3)]. The seven-term model matches the sparsity pattern of FHN while the six-term model is equivalent under the shift  $w \mapsto w  - I$ in Eq.~\eqref{eq:fhn}. \rv{Indeed by taking the true FHN model in Eq.~\eqref{eq:fhn}, solving for $w$ in terms of $v, \dot{v}$ and substituting into the $\dot{w}$ equation, we obtain a second-order \textit{reduced model} solely in $v$. Performing a similar reduction of our learned six-term model (SI) we see it has the same structure and similar coefficients as the true FHN model [Fig.~\ref{fig:fig1_fhn}(b)], confirming that HDI has recovered a two-variable model that is equivalent to ground-truth FHN.} We develop an algebraic procedure (SI) to automatically verify these polynomial model reductions on future examples.

\par
At this point, one might hope to avoid using hidden variables and their associated ambiguities by learning the reduced higher-order equation in the observed variable directly~\cite{lainscsek2003global, somacal2020uncovering}. However, even simple multivariate systems can give rise to complex reduced higher-order equations that are often less sparse, implicit and contain fractional powers~\cite{gouesbet1994global, gottlieb1996question, sprott1997simplest, linz1997nonlinear, eichhorn1998transformations, eichhorn1999classes, eichhorn2002simple, letellier2005relation, mendes2021diffeomorphical}; for example $\dot{x} = xy^3, \, \dot{y} = x$  reduces to $ x\ddot{x} = \dot{x}^2 + 3x^{7/3} \dot{x}^{2/3}$. Working with reduced-models directly would require learning dense implicit ODEs with more candidate terms~\cite{mangan2016inferring, kaheman2020sindy}, a challenging approach which can be ill-posed~\cite{kunkel2006differential}. A general advantage of \lq first-order\rq{} HDI is that it robustly learns multivariate explicit ODE models that are sparse and integrable, avoiding the above complications. We next apply HDI to identify quantitative models from experimental data for neuron activity and chemical reactions. 

\par
Figure~\ref{fig:fig2_neuron}(a) shows  experimental measurements~\cite{paydarfar2006noisy, physionet2000} of the membrane potential $v$ in the giant axon of the North Atlantic longfin inshore squid (\textit{Loligo pealeii}) in response to noisy stimulus input currents. Following previous spike train model formulations~\cite{fitzhugh1961impulses, morris1981voltage, hodgkin1952currents, gerstner2014neuronal}, we apply HDI to the time series data for $v$ to learn a sparse two-variable model [Fig.~\ref{fig:fig2_neuron}(a,b)].
Consistent with prior descriptions of neuron dynamics~\cite{IZH07}, the phase portrait of the discovered seven-term model is governed by a homoclinic orbit [Fig.~\ref{fig:fig2_neuron}(c)]. Importantly, the model generalizes to describe recordings from different squids, yielding consistent  coefficients across all samples [Fig.~\ref{fig:fig2_neuron}(d,e)].
\par

For a second more challenging HDI application,  we performed  Belousov-Zhabotinsky (BZ) reaction experiments~\cite{EPS98}. Over the course of the reaction a substrate species is slowly consumed that fuels the periodic rise and decay of intermediary reagents far from thermodynamic equilibrium. The basic reaction scheme~\cite{TAY02} involves more than 20 chemical species and 40 reaction steps. A plethora of different chemical models have been developed that capture the BZ reaction qualitatively~\cite{FIE72b,TYS82,ROV84a,ZHA93c,REN15a}.
In our experiments, the repeated  oxidation and reduction of the metal catalyst ferroin produces a periodic change in color of the solution from red to light blue~[Fig.~\ref{fig:fig3_bz}(a)]. The recorded average color of the solution follows  a 1D curve in color space which we map to our single observed coordinate $c(t)$ (SI). Working with polynomial approximations consistent with established two-variable BZ models~\cite{EPS98}, we optimize over all two-variable ODEs that are linear in the $\dot{c}$ equation and quartic in the $\dot{h}$ equation. Using this library, HDI discovers a seven-term model that accurately fits the color dynamics~$c(t)$ for BZ reactions \rv{[Fig.~\ref{fig:fig3_bz}(a, d)]} with \rv{parameters that vary smoothly across the different reactant concentrations in each experiment [Fig.~\ref{fig:fig3_bz}(e)].}
Furthermore, the phase portrait  of the learned model correctly captures the dynamical properties of the BZ reaction~\cite{EPS98}, showing an unstable fixed point enclosed in a stable limit cycle with a typical cubic-shaped nullcline $\dot{h} = 0$ [Fig.~\ref{fig:fig3_bz}(c)].

\par
\begin{figure}
    \centering
    \includegraphics{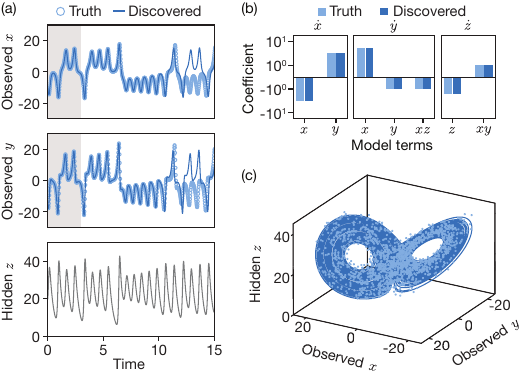}
    \caption{HDI discovers true Lorenz system from observations of $x$ and $y$ coordinates. (a) Given observations of only the $x$ and $y$ coordinates (gray region), the learned model predicts the evolution for several additional Lyapunov time-scales. (b) Lorenz model terms and coefficients are discovered exactly by HDI search solely from data in gray region of previous panel. (c) Reconstructed attractor of learned model closely agrees with the true Lorenz attractor (simulation parameters and noise robustness in SI).}
    \label{fig:fig4_lorenz}
\end{figure}

HDI straightforwardly extends to higher-dimensional nonlinear systems. For example, when only given observations of the $x$ and $y$ coordinates of the 3D Lorenz system for one or two lobe transitions [gray-shaded in Fig.~\ref{fig:fig4_lorenz}(a)], a HDI search over all polynomial three-variable ODEs in $(x, y, z)$ with quadratic interactions recovers the exact Lorenz equations with correct coefficient values (modulo a trivial scaling of the hidden $z$ variable) [Fig.~\ref{fig:fig4_lorenz}(b); SI]. The learned model has the correct attractor dynamics and can predict the $x, y,$ and $z$ dynamics substantially beyond the training interval  [Fig.~\ref{fig:fig4_lorenz}(a,c)]. Further analysis shows that, even when only given observations of $x$,  HDI learns a predictive model for Lorenz dynamics, albeit with reduced predictive power~(SI).

\par

To conclude, by combining sensitivity methods and ranked choice voting, HDI can discover parsimonious predictive models from partial noisy observations of oscillatory and chaotic dynamics  without extensive preprocessing of time-series data.  The above framework can be directly applied to experimental observations of biophysical, ecological and other systems, for which ODE models can inform the prediction, control and optimal perturbations~\cite{stepaniants2020inferring} of dynamical behavior. By mapping time series to ODE model coefficients, HDI can help facilitate clustering of dynamical  data, as those appearing in health ~\cite{UMETANI1998} and climate~\cite{Vettoretti2022} studies. 

%TC:ignore

\begin{acknowledgements}
\rv{All source code is available at \url{https://github.com/adh18/HiddenDynamicsInference}.} G.S. and A.D.H. contributed equally and are joint first authors. We thank Keaton Burns and Peter Baddoo for helpful discussions on partially observed systems. We acknowledge the MIT SuperCloud and Lincoln Laboratory Supercomputing Center~\cite{reuther2018interactive} for providing HPC resources. J.F.T. acknowledges support through a Feodor Lynen Fellowship of the Alexander von Humboldt Foundation. G.S. acknowledges support through a National Science Foundation Graduate Research Fellowship under Grant No. 1745302. This work was supported by a MathWorks Science Fellowship (A.D.H.), Sloan Foundation Grant G-2021-16758 (J.D.), and the Robert E. Collins Distinguished Scholarship Fund (J.D.).
\end{acknowledgements}

\nocite{HomotopyContinuation.jl}

\bibliographystyle{apsrev4-2}
\bibliography{bibliography}

%\detailtexcount{main}

%TC:endignore
\end{document}

\title{Supplementary Information:~\\
Discovering dynamics and parameters of nonlinear oscillatory and chaotic systems
from partial observations}

\author{George Stepaniants}
\affiliation{Department of Mathematics, Massachusetts Institute of Technology, Cambridge, Massachusetts 02139-4307, USA}
\author{Alasdair D. Hastewell}
\affiliation{Department of Mathematics, Massachusetts Institute of Technology, Cambridge, Massachusetts 02139-4307, USA}
\author{Dominic J. Skinner}
\affiliation{Department of Mathematics, Massachusetts Institute of Technology, Cambridge, Massachusetts 02139-4307, USA}
\affiliation{NSF-Simons Center for Quantitative Biology, Northwestern University, Evanston, Illinois 60208, USA}
\author{Jan F. Totz}
\affiliation{Department of Mathematics, Massachusetts Institute of Technology, Cambridge, Massachusetts 02139-4307, USA}
\author{J\"orn Dunkel}
\affiliation{Department of Mathematics, Massachusetts Institute of Technology, Cambridge, Massachusetts 02139-4307, USA}

\date{\today}

\maketitle

\onecolumngrid

\rv{\section{Comparison to existing hidden variable model inference approaches}}

\rv{In canonical form, data-driven dynamics discovery consists of fitting the parameters $\mathbf{p}$ of an ordinary differential equation  (ODE) $\dot{x} = f(x, \mathbf{p})$ to a given dataset. Serious consideration has been given to this problem ever since early computers were used to numerically solve differential equations~\cite{BoxCoutie}. Renewed interest in the problem came with increasing computational power in the 1980s and 1990s~\cite{aguirre2009modeling}. With the increasing prevalence of machine learning techniques and the explosion of high resolution data acquisition, once again, attention has turned to this problem in recent years~\cite{rackauckas2020universal, bakarji2022discovering, ribera2022model, raissi2020hidden, heinonen2018learning}. Here we review previous approaches to ODE  model inference focusing particularly on those techniques that can be extended to partial observations and contrasting them with the approach outlined in the main text and explained in more detail in later sections.
}

\rv{
\subsection{Parametric models}
The simplest case is when the form of the ODE right hand side $f(x, \mathbf{p})$ is fixed ahead of time by some knowledge of the system, for example the FitzHugh-Nagumo model with unkown parameters $\textbf{p}= (I,a,b,\tau)$. Frequently $\mathbf{p}$ will be low dimensional, consisting of only a few parameters. Fitting these parameters can be formulated as an optimization problem minimizing a data loss $\sum_i |\dot{x}_i - f(x_i, \mathbf{p})|^2$. Approximations of the time derivative can be calculated by finite differences, smooth derivative approximations~\cite{breugel9241009, varah1982spline, aguirre2009modeling} or weak formulations~\cite{messenger2021weak}. Minimizing the data loss to find $\textbf{p}$ can be done using either linear~\cite{BROWN199471, brunton2016discovering} or nonlinear~\cite{varah1982spline, dattner2020separable} least squares based on the structure of the dynamical equations. For instance, the parameters of the Lorenz system enter linearly in the right hand side while the parameters of the nonlinear pendulum $\ddot{x} = -A\sin(\omega x)$ do not. These regression approaches, however, cannot be used for partial observations, since some coordinates of the system are not available to compute time derivatives. Furthermore, such methods are not robust and can overfit noisy data, as they impose no causal constraints that the observed data at time $t + \Delta t$ is related to the data at time $t$ through the evolution of an underlying differential equation.}

\rv{
\subsection{Optimization methods}
In the case of partial observations or noise corrupted data, a causal formulation of the problem is needed, which encodes prior knowledge that observed data are related through the evolution of a differential equation. When the integrated solution to the ODE model has a closed form such as with linear time-invariant systems, least squares fitting can be used to match the integrated solution to the data~\cite{cohen2023schrodinger}. For more complicated models, we need to use numerical approximations to the integrated ODE solution and fit the parameters by solving suitably chosen often non-linear and non-convex optimization problems which depend only on the observed variables. The causal nature of the problem can be encoded by adding additional penalty terms to the data loss enforcing an ODE structure, leading to a large class of approaches known as data assimilation methods~\cite{ribera2022model, law2015data}. These approaches typically enforce that the fit dynamics is consistent with a single time step of an ODE model although generalizations to $n$ steps have shown better stability and noise robustness~\cite{rudy2019deep, fullana1997parameter, abarbanel1990prediction}. While better suited for data with partial observations and noise, there is still no guarantee that the integrated model is stable over long times or fits the data when these finite steps are performed sequentially. A stronger constraint on the ODE structure is that the solution $\{x(t,\textbf{p})\}$ comes from numerically integrating the model, which is then fitted to the data. We can achieve this hard constraint by directly optimizing through the numerical ODE solver. Such optimization techniques roughly fall into two categories: (1) gradient free methods like root finding~\cite{breeden1990reconstructing}, multiple shooting~\cite{baake1992fitting}, Nelder-Mead, particle swarm optimization~\cite{krueger2017parameter}, or Kalman filters~\cite{aguirre2005using, levine2022framework, schneider2022ensemble}, and (2) gradient based methods like Newton's method, ADAM~\cite{kingma2014adam}, and BFGS~\cite{liu1989limited} where derivatives are approximated either using finite differencing on the ODE solutions, sensitivity methods or automatic differentiation~\cite{cao2003adjoint, rackauckas2020universal}. These techniques have been applied successfully to learn periodic, chaotic, and transient dynamics~\cite{rackauckas2020universal}. The question of parameter identifiability, which studies whether model coefficients can be determined uniquely from the data, has also been extensively explored for parametric models~\cite{villaverde2016structural, massonis2023distilling}. However, parametric models are only applicable when we have some prior knowledge about the form of the model governing the dynamics in our data, which is frequently not the case for experimental systems.}

\rv{
\subsection{Nonparametric models}
If we cannot assume a particular parametric form of the equations, a more expressive and less constrained right hand side function $f$ is needed. Taking $f$ to be a general function with a large set of model parameters, for instance a neural network, results in a nonparametric model. It is called nonparametric as the large set of parameters has no inherent interpretation. Examples of such functions $f$ that have been applied to dynamical inference include locally linear models~\cite{farmer1987predicting, costa2019}, bases expansions~\cite{crutchfield1987equations} (sometimes referred to as atlases or libraries including polynomial~\cite{brunton2016discovering}, rational polynomial~\cite{correa2000modeling, mangan2016inferring}, wavelet~\cite{wei2004identification}, and radial basis functions~\cite{suontausta1994modelling}), discrete-time perceptrons~\cite{albano1992using, bakker2000learning}, and neural networks (referred to as neural ODEs)~\cite{rackauckas2020universal}. The techniques for fitting parametric models outlined above can also be applied to these general functions. These techniques often discretize the ODE during fitting; directly learning a discretized form of the dynamics $x_{t+\Delta t} = f(x_{t}, \mathbf{p})$, is called a discrete nonparametric model~\cite{crutchfield1987equations, suontausta1994modelling, correa2000modeling, albano1992using, bakker2000learning}.}

\rv{
\subsection{Model selection}
Without prior knowledge about the form of the dynamics, nonparametric models can fit an ODE to data. Yet the resulting model lacks physical interpretability. This drawback prompted the development of automated methods for learning physically interpretable ODE models, without assuming prior knowledge of the dynamics.  A popular approach is to first learn a nonparametric model given as a linear combination of basis terms (i.e. atlas, library), and to subsequently perform model selection keeping only a few basis terms in the expansion~\cite{crutchfield1987equations, brunton2016discovering}. Various techniques for basis term selection include sparse identification~\cite{brunton2016discovering}, bootstrapping~\cite{fasel2022ensemble}, information criteria (MLE, AIC, BIC, MDL) ranking~\cite{crutchfield1987equations, small2002minimum, dong2023improved}, and hypothesis testing~\cite{wang2020perspective, wang2019variational}. When partial knowledge of the ODE model is available, a parametric interpretable ODE model can be added to a nonparametric model and jointly fit to the data, an approach called hybrid modeling~\cite{levine2022framework}.
}

\rv{
\subsection{Reduced higher-order models}
When some coordinates of the system are not observed, one approach is to learn a higher-order model in the coordinates which can be observed. We call such a model a \textit{reduced model}. This can be done directly by taking continuous derivatives~\cite{packard1980geometry} of the observed coordinates~\cite{casdagli1991state} or indirectly by constructing a state vector from successive time lags~\cite{bakarji2022discovering, sugihara1990nonlinear, sugihara2012detecting}. However, even simple multivariate systems give rise to complicated reduced higher-order equations which often have many terms, are implicit, and contain fractional powers~\cite{gouesbet1994global, gottlieb1996question, sprott1997simplest, linz1997nonlinear, eichhorn1998transformations, eichhorn1999classes, eichhorn2002simple, letellier2005relation, mendes2021diffeomorphical}. Learning reduced models directly requires learning dense implicit ODEs with many candidate terms~\cite{mangan2016inferring, kaheman2020sindy}, a challenging approach which can be ill-posed~\cite{kunkel2006differential}. Derivative fitting methods~\cite{gouesbet1994global} require computing higher order numerical derivatives of observed coordinates which is prohibitive for noisy or highly oscillatory data. Gradient free or gradient based methods also face serious problems due to the appearance of rational terms which complicate the optimization as the dynamics of the learned model can frequently set the denominator of these terms close to zero. While reduced models for partially observed systems are a useful tool for model comparison, as shown throughout this paper, they are difficult to directly learn from data.}

\rv{
\subsection{Our contribution}
Our HDI methodology falls into the class of physically interpretable methods, and learns ODE models as a sparse combination of polynomial or trigonometric basis functions. However, HDI uses a novel robust procedure for model selection by sampling the space of possible ODE models, forming a cluster of the best fit models, and keeping those basis terms in our final model that have the least variation in their coefficients across all model fits. This allows HDI to robustly select a few important model terms from libraries containing on the order of 10-100 candidate terms, thus exploring a broad range of potential ODE models. HDI's core model fitting procedure uses gradient based sensitivity methods which are robust to high levels of noise in the data, ensure the learned models accurately capture the experimental dynamics, and are stable over long-time integration. All source code for our HDI methodology and instructive examples are available at \url{https://github.com/adh18/HiddenDynamicsInference}.
}

\section{Reductions of dynamical systems to higher order models}

Whilst we are unable to compare hidden variables of a model to the unobserved coordinates of a partially observed system, we can reduce a model with hidden variables into a higher-order ODE model where only the observed coordinates are present.
Comparing the terms and coefficients of learned models in this ``reduced" model space allows us to deduce whether two models are similar in their description of the dynamics of the observed coordinates. We demonstrate these ideas on the motivating example of the FitzHugh-Nagumo system and the closely related van der Pol oscillator.

\subsection{FitzHugh-Nagumo model}
The FitzHugh-Nagumo model is a two-dimensional system in $(v, w)$ where $v$ simulates the membrane potential (voltage) of a neuron and $w$ is a slow recovery variable modeling the potassium ion ($K^+$) current. It is given by a nonlinear ODE of the form
\begin{subequations} \label{eq:fhn}
\begin{align}
\dot{v} = v - cv^3 - w + I \label{eq:fhnv}\\
\dot{w} = \frac{1}{\tau}\Big(v + a - bw\Big) \label{eq:fhnw}
\end{align}
\end{subequations}
which is a simplified 2D version of the Hodgkin-Huxley model.

Suppose we are only allowed to observe the membrane potential $v$ of this system, as is commonly the case in practice. Then it is possible to write out a differential equation that solely depend on $v$ and its derivatives. To see this, we can solve for $w$ in~\eqref{eq:fhnv} to get that
\begin{equation}
    w = -\dot{v} + v - cv^3 + I
\end{equation}
which can be substituted into~\eqref{eq:fhnw} to obtain
\begin{equation}
    \ddot{v} - \dot{v} + 3cv^2\dot{v} = -\frac{1}{\tau}v - \frac{a}{\tau} + \frac{b}{\tau}\Big(-\dot{v} + v - cv^3 + I\Big)
\end{equation}
that can be expanded into
\begin{equation}\label{eq:redfhnv}
    \ddot{v} = -3cv^2\dot{v} + \Big(1 - \frac{b}{\tau}\Big)\dot{v} - \frac{bc}{\tau}v^3 + \frac{b-1}{\tau}v + \frac{bI - a}{\tau}.
\end{equation}
We say that equation~\eqref{eq:redfhnv} is in ``reduced form" because it depends solely on the value and derivatives of the observable $v$.

 Similarly, we can perform the same exercise and find a reduced model that describes the dynamics of the $w$ coordinate. First we solve for $v$ in~\eqref{eq:fhnw} to get that
\begin{equation}
    v = \tau\dot{w} + bw - a.
\end{equation}
Substituting this into~\eqref{eq:fhnv} gives us
\begin{equation}
    \ddot{w} = \frac{1}{\tau}\Big[(\tau-b)\dot{w} + (b-1)w - a + I - c(\tau\dot{w} + bw - a)^3\Big].
\end{equation}
which can be expanded out to
\begin{equation}
\begin{aligned}
    \ddot{w} &= -c\tau^2\dot{w}^3 - 3bc\tau\dot{w}^2w + 3ac\tau\dot{w}^2 - b^2c\dot{w}w^2 + 2ab\dot{w}w + \Big(1 - 3a^2c - \frac{b}{\tau}\Big)\dot{w}\\
    &\quad- \frac{b^3c}{\tau}w^3 + \frac{3ab^2c}{\tau}w^2 + \frac{-3a^2bc + b - 1}{\tau}w + \frac{I - a + a^3c}{\tau}.
\end{aligned}
\end{equation}

Naturally we may ask if there exist other dynamical systems in $(v, w)$ which have the exact same dynamics in one of their coordinates. This turns out to be trivially true, as any model of the form
\begin{equation}
\begin{aligned}
\dot{v} &= v - cv^3 - \frac{1}{\alpha}w + I +\frac{\beta}{\alpha}\\
\dot{w} &= \frac{\alpha}{\tau}\Big(v + a + \frac{b\beta}{\alpha} - \frac{b}{\alpha} w\Big)
\end{aligned}
\end{equation}
with $v$ observed and $w$ unobserved reduces to the same reduced model~\eqref{eq:redfhnv} in $v$ for any $\alpha, \beta \in \mathbb{R}$ with $\alpha \neq 0$. In fact, the model above is related to the original FHN equations~\eqref{eq:fhn} through the linear transformation $w \mapsto \alpha w + \beta$.

Furthermore, the parameters $\alpha, \beta$ in the linear transformation above can be chosen carefully so as to change the sparsity pattern of the original FHN model (e.g. change the polynomial terms appearing on the right hand side). For example, we can set $\beta = -\alpha I$ for any $\alpha \neq 0$ and this will remove the constant/bias term in the FHN equation for $\dot{v}$. Alternatively, we can set $\beta = -a\alpha/b$ for any $\alpha \neq 0$ which will remove the constant term in the FHN equation for $\dot{w}$. Hence, given dynamics from a partially observed ground-truth system, there can exist sparser models (containing fewer terms on the right hand side) that exactly reproduce the dynamics of its observed coordinates.\\

From now on, we refer to transformations of hidden (unobserved) variables that preserve the dynamics of observed variables as \textit{model symmetries}. The class of model symmetries is much larger than the set of linear transformations of hidden variables. In general, any diffeomorphic transformation (differentiable map whose inverse is also differentiable) that preserves the observed coordinates constitutes a model symmetry. On the example of FHN with $v$ observed, all diffeomorphisms $\Phi: \mathbb{R}^2 \to \mathbb{R}^2$ that preserve the observed coordinate take the form
\begin{equation}
    \Phi\begin{pmatrix}v\\ w\end{pmatrix} = \begin{pmatrix}v\\ f(v, w)\end{pmatrix}
\end{equation}
where $f: \mathbb{R}^2 \to \mathbb{R}$ and $\partial_w f$ is nonzero everywhere. This ensures by the implicit function theorem that there exists (at least locally) an inverse function $f^{-1}: \mathbb{R}^2 \to \mathbb{R}$ such that $f(v, f^{-1}(v, w)) = f^{-1}(v, f(v, w)) = w$.

Using these facts, from the original FHN equations, we can transform $(v, w) \mapsto f(v, w)$ to get the new system
\begin{equation}\label{eq:transformed}
\begin{aligned}
\dot{v} &= v - cv^3 - f^{-1}(v, w) + I\\
\dot{w} &= \partial_vf(v, f^{-1}(v, w))\Big(v - cv^3 - f^{-1}(v, w) + I\Big) + \frac{1}{\tau}\partial_wf(v, f^{-1}(v, w))\Big(v + a - bf^{-1}(v, w)\Big).
\end{aligned}
\end{equation}
By definition, this new model still reduces to the exact same higher-order system~\eqref{eq:redfhnv} in $v$. We may further ask which diffeomorphic transformations (model symmetries) $\Phi$ lead to new polynomial models with a small number of terms on the right hand side (e.g. parsimonious polynomial models). In general, for a polynomial ODE model with several observed variables $\mathbf{x} = (x_1, \hdots, x_m) \in \mathbb{R}^m$ and hidden variables $\mathbf{h} = (h_{m+1}, \hdots, h_M) \in \mathbb{R}^{M-m}$, all model symmetries (with rare exceptions) take the form
\begin{equation}
    \Phi\begin{pmatrix}\mathbf{x}\\ \mathbf{h}\end{pmatrix} = \begin{pmatrix}\mathbf{x}\\ \mathbf{b} + \mathbf{A}\mathbf{h} + \mathbf{P}(\mathbf{x})\end{pmatrix}
\end{equation}
where $\mathbf{b} \in \mathbb{R}^{M-m}$ and $\mathbf{A} \in \mathbb{R}^{(M-m) \times (M-m)}$ linearly transform the hidden variables and $\mathbf{P}(\mathbf{x}) = (p_1(\mathbf{x}), \hdots, p_{M-m}(\mathbf{x}))^T$ is a vector of polynomial functions of the observed variables $\mathbf{x}$. Other model symmetries $\Phi$ which preserve the polynomial structure of an ODE may contain rational polynomial terms but appear quite rarely in practice.

Continuing with our FHN example, we investigate this class of diffeomorphisms \begin{equation}\label{eq:ex_transform}
    f(v, w) = \alpha + \beta w + p(v)
\end{equation}
with inverse
\begin{equation}
    f^{-1}(v, w) = \frac{w - \alpha - p(v)}{\beta}.
\end{equation}
Substituting these expressions for $f(v, w)$ and $f^{-1}(v, w)$ into~\eqref{eq:transformed} we get
\begin{equation}
\begin{aligned}
\dot{v} &= v - \frac{1}{\beta}w - cv^3 + \frac{1}{\beta}p(v) + \Big(\frac{\alpha}{\beta} + I\Big)\\
\dot{w} &= \frac{a\beta + \alpha b}{\tau} + \frac{\beta}{\tau}v + \frac{b}{\tau}p(v) + (\frac{\alpha}{\beta} + I)p'(v) + vp'(v) - cv^3p'(v) + \frac{1}{\beta}p(v)p'(v) - \Big(\frac{b}{\tau} + \frac{1}{\beta}p'(v)\Big)w.
\end{aligned}
\end{equation}
By choosing $p(v) = v$ we get
\begin{equation}
\begin{aligned}
\dot{v} &= \Big(1 + \frac{1}{\beta}\Big)v - \frac{1}{\beta}w - cv^3 + \Big(\frac{\alpha}{\beta} + I\Big)\\
\dot{w} &= \Big(\frac{a\beta + \alpha b}{\tau} + \frac{\alpha}{\beta} + I\Big) + \Big(\frac{b + \beta}{\tau} + \frac{1}{\beta} + 1\Big)v - cv^3 - \Big(\frac{b}{\tau} + \frac{1}{\beta}\Big)w
\end{aligned}
\end{equation}
and setting $\alpha, \beta$ appropriately to remove the constant and linear terms in $v$ in the equation for $\dot{w}$ gives us
\begin{equation}
\begin{aligned}
\dot{v} &= \Big(1 + \frac{1}{\beta}\Big)v - \frac{1}{\beta}w - cv^3 + \Big(\frac{\alpha}{\beta} + I\Big)\\
\dot{w} &= -cv^3 - \Big(\frac{b}{\tau} + \frac{1}{\beta}\Big)w.
\end{aligned}
\end{equation}
which is equivalent to the original FHN system from the perspective of the $v$ coordinate.

\subsection{Van der Pol oscillator}
An even simpler model of a nonlinear oscillator is the Van der Pol oscillator given by the equations
\begin{subequations} \label{eq:vdp}
\begin{align}
\dot{x} &= \mu(x - cx^3 - y)\label{eq:vdpx}\\
\dot{y} &= \frac{1}{\mu}x\label{eq:vdpy}
\end{align}
\end{subequations}
which becomes the simple harmonic oscillator for $\mu \to 0$ and for $\mu > 0$ it exhibits a relaxation oscillation around a stable limit cycle. It differs from the FHN system~\eqref{eq:fhn} because it has no constant terms and the second variable $y$ does not force itself in the $\dot{y}$ equation. The reduced forms of this equation are trivial to derive and are given by
\begin{equation}
    \ddot{x} = -3c\mu x^2\dot{x} + \mu\dot{x} - x
\end{equation}
if $x$ is observed and similarly
\begin{equation}
    \ddot{y} = -c\mu^3\dot{y}^3 + \mu\dot{y} - y
\end{equation}
if $y$ is observed. Note that the reduced equation in $x$ above is actually the original second-order differential equation for the Van der Pol oscillator.

A further variant of this system is the \textit{constantly forced} Van der Pol oscillator given by
\begin{subequations} \label{eq:vdp_cf}
\begin{align}
\dot{x} &= \mu(x - cx^3 - y)\label{eq:vdpx_cf}\\
\dot{y} &= \frac{1}{\mu}(a - x)\label{eq:vdpy_cf} 
\end{align}
\end{subequations}
where the additional constant forcing $a$ leads to bifurcations in the model dynamics. For $|a| < 1/\sqrt{3c}$ the system exhibits a relaxation oscillation near a stable limit cycle as in the prior model. However, for $|a|$ slightly larger than $1/\sqrt{3c}$ the constantly forced Van der Pol system exhibits excitations, long excursions through phase space before returning to a globally attractive fixed point (see Strogatz~\cite{strogatz2018nonlinear}). Excitations are an important property of dynamical models of neurons, and such behaviors cannot be achieved by the unforced Van der Pol model.

The FHN model analyzed in the previous section similarly exhibits bifurcations in its dynamics and can display periodic oscillations as well as excitations depending on the value of its bias parameter $a$. Remarkably, our model search procedure described in the following section can occasionally learn the constantly forced (excitable) Van der Pol oscillator when given noisy simulated data from the FHN system. This implies that the constantly forced Van der Pol oscillator is an even simpler (sparser) polynomial model whose dynamics is similar to the FHN system.

\subsection{Lotka-Volterra}
The Lotka-Volterra is a model for predator-prey interactions which exhibits limit cycle dynamics and has a Hamiltonian structure. Here we reduce the two-species Lotka-Volterra model
\begin{subequations} \label{eq:lv}
\begin{align}
\dot{x} &= \alpha x - \beta xy \label{eq:lvx}\\
\dot{y} &= \delta xy - \gamma y \label{eq:lvy}
\end{align}
\end{subequations}
in both coordinates. To reduce in $x$, we first use~\eqref{eq:lvx} to write
\begin{equation}
    y = -\frac{\dot{x} - \alpha x}{\beta x} = -\frac{1}{\beta}\frac{\dot{x}}{x} + \frac{\alpha}{\beta}.
\end{equation}
Differentiating once we get that
\begin{equation}
    \dot{y} = -\frac{1}{\beta}\frac{\ddot{x}}{x} + \frac{1}{\beta}\frac{\dot{x}^2}{x^2}
\end{equation}
and substituting $y, \dot{y}$ into~\eqref{eq:lvy} we get that
\begin{equation}
    -\frac{1}{\beta}\frac{\ddot{x}}{x} + \frac{1}{\beta}\frac{\dot{x}^2}{x^2} = -\frac{\delta}{\beta}\dot{x} + \frac{\alpha\delta}{\beta}x + \frac{\gamma}{\beta}\frac{\dot{x}}{x} - \frac{\alpha\gamma}{\beta}
\end{equation}
and multiplying by $\beta x^2$ on both sides gives us
\begin{equation}
    -x\ddot{x} + \dot{x}^2 = -\delta x^2\dot{x} + \alpha\delta x^3 + \gamma x\dot{x} - \alpha\gamma x^2
\end{equation}
so we can finally write the reduced equation in $x$ as
\begin{equation}
    x\ddot{x} - \dot{x}^2 - \delta x^2\dot{x} + \gamma x\dot{x} + \alpha\delta x^3 - \alpha\gamma x^2.
\end{equation}
By symmetry, taking $\alpha \mapsto -\gamma, \beta \mapsto -\delta, \gamma \mapsto -\alpha$ and $\delta \mapsto -\beta$ we can write the reduced equation in $y$ as
\begin{equation}
    \dot{y}\ddot{y} - y\ddot{y} - \gamma\dot{y}^2 + \beta y^2\dot{y} + (\alpha - \gamma)y\dot{y} + \gamma\beta y^3 - \alpha\gamma y^2.
\end{equation}

\rv{
\section{Gaussian Additive Noise}
Here we describe how noise is added to the simulated dynamical system trajectories when testing the noise robustness of our HDI method. Recall that given $m$ observed coordinates $x_1, \hdots, x_m$ of a simulated dynamical system, we solve a corresponding ODE to generate the trajectory data $x_k(t_i)$ for coordinates $1 \leq k \leq m$ sampled at timepoints $t_i$ where $1 \leq i \leq n$. From this, we now generate noisy observations by adding $p\%$ relative Gaussian noise as
\begin{equation}
    y_{ik} = x_k(t_i) + \frac{p}{100} \cdot \text{std}(\{x_k(t_i)\}_{i=1}^n) \cdot \eta_{ik}
\end{equation}
where $\eta_{ik} \sim \mathcal{N}(0, 1)$ are independent identically distributed Gaussian random variables and
\begin{equation}
    \text{std}(\{z_i\}_{i=1}^n) = \sqrt{\frac{1}{n}\sum_{i=1}^n\Big(z_i - \frac{1}{n}\sum_{i=1}^n z_i\Big)^2}
\end{equation}
is the standard deviation of $n$ samples.
}

\section{Model Search Pipeline}\label{sec:model_search}
Here we describe our general pipeline which enables us to discover hidden variable ODE models for the FitzHugh-Nagumo (FHN) relaxation oscillator and chaotic Lorenz system as well as real experimental recordings from squid axons and the Belousov-Zhabotinsky (BZ) chemical reaction.

Our aim is to discover ODE models consisting of $M$ scalar variables, such that the first $m$ variables $\mathbf{x}(t) = \{x_k(t)\}_{k=1}^m$ fit observed data $\{(t_i, \mathbf{y}_i)\}_{i=1}^n$ where $\mathbf{y}_i = \{y_{ik}\}_{k=1}^m$ and the next $M - m$ variables $\{h_k(t)\}_{k=m+1}^M$ are hidden.
Given a set of $m$ time series observations, there are often several sparse hidden variable models that can closely fit the observed dynamics. We develop a general model search/sweep procedure for selecting the best hidden variable model among these choices. Our procedure involves training many HDI models, separating them into different model clusters based on their coefficients, selecting the HDI model type which corresponds to the largest cluster, and finally sparsifying this chosen model by removing those terms whose coefficients are small or highly variable.

All ODE models learned in this paper result from the model sweep procedure outlined below:\\

\begin{figure}
    \centering\includegraphics{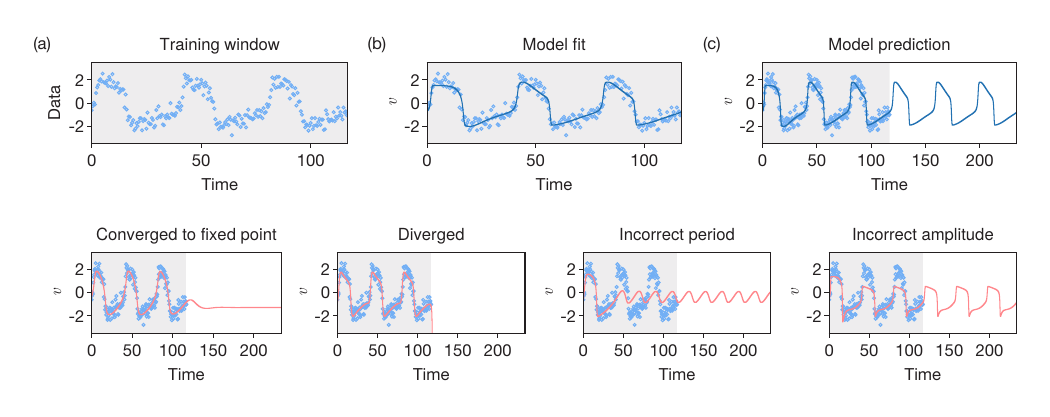}
    \caption{Three steps for testing a HDI model fit to data. (a) First we are given \rv{training data} which \rv{consists of three periods} of noisy oscillations of the FHN $v$-coordinate corrupted by 30\% Gaussian noise. (b) Example HDI ODE model is optimized to fit \rv{training} data from $v$-coordinate. (c) Learned HDI model is simulated past the \rv{training} data window and we verify that it continues to behave periodically. Bottom row shows possible failure cases, namely a learned HDI model may converge to a fixed point, diverge to infinity, or learn a periodic dynamics with the incorrect period or amplitude. All such problematic models are removed from the HDI model search procedure.}
    \label{fig:fhn_fit_types}
\end{figure}

\par\textbf{Initial Optimization}
\begin{enumerate}
    \item Given a recorded time series, we select a window of training data points $\{(t_i, \mathbf{y}_i)\}_{i=1}^n$ where $t_1 < t_2 < \hdots < t_n$. If the time series dynamics behaves periodically, then we choose our training window $[t_1, t_n]$ to contain three periods of oscillation. This heuristic ensures that our optimization converges to a model with periodic behavior from most random initializations. Choosing too few periods in the training window results in learned models with only transient periodic behavior which then either diverge or converge to a fixed point. On the other hand, choosing too many time periods (or generally a large \rv{training} window) leads the HDI optimization to ODE models that converge to a fixed point close to the mean of the data, a common feature in sensitivity-based ODE inference methods. If the time series dynamics behaves chaotically, as with the Lorenz system, we choose our training window $[t_1, t_n]$ to contain two or three branch switches which helps our algorithm converge to a chaotic model more frequently. Providing too few branch switches leads to models that eventually fall on a periodic limit cycle. As before, we do not include more than three branch switches as long time windows lead to models which converge to fixed points.
    
    \item We define an $M$-variable polynomial HDI model with degree combination $(d_1, \hdots, d_M)$ as
    \begin{subequations} \label{eq:hfm}
    \begin{align}
        \dot{x}_k &= \tau_k\sum_{|\bm\alpha| \leq d_k}c_{\bm\alpha}^k x_1^{\alpha_1} \hdots h_M^{\alpha_M}, \quad x_k(0) = b_k, \quad 1 \leq k \leq m\\
        \dot{h}_k &= \tau_k\sum_{|\bm\alpha| \leq d_k} c_{\bm\alpha}^k x_1^{\alpha_1} \hdots h_M^{\alpha_M}, \quad h_k(0) = b_k, \quad m < k \leq M
    \end{align}
    \end{subequations}
    where the free parameters are the initial conditions $b_k$, polynomial coefficient vectors $\mathbf{c}_k = \{c_{\bm\alpha}^k\}_{|\bm\alpha| \leq d_k}$, and time scales $\tau_k$ for all $1 \leq k \leq M$. By definition, the degree of any polynomial term in equation $k$ does not exceed $d_k$. The power of each monomial term above is expressed in multi-index notation $\bm\alpha = (\alpha_1, \alpha_2, \hdots, \alpha_M)$ where each entry $\alpha_r \in \mathbb{N}$ denotes the power of $x_r$ and we write $|\bm\alpha| = \sum_{k=1}^M \alpha_k$ to denote the total degree of the monomial term.

    In a model sweep, each HDI model is initialized randomly by setting the initial conditions of its observed variables to $b_k = y_{1k}$ for $1 \leq k \leq m$ and of its hidden variables to $b_k = 1$ for $m+1 \leq k \leq M$. Lastly, we choose the $M$ coefficient vectors $\{\mathbf{c}_k\}_{k=1}^M$ independently at random as uniformly distributed vectors on the unit sphere and set the time scales of all equations to $\tau_k = 0$ for $1 \leq k \leq M$. 
    
    We take 11 choices of sparsity penalties $\lambda \in \{10^{-5}, 5\times10^{-5}, 10^{-4}, 5\times10^{-4}, 10^{-3}, 5\times10^{-3}, 10^{-2}, 5\times10^{-2}, 10^{-1}, 5\times10^{-1}, 10^{0}\}$, and for each choice we randomly initialize 200 HDI models. Here the number of random initializations was chosen large enough to sample the model space. The range of sparsity penalties is chosen such that the magnitude of the penalty term is less than or equal to the magnitude of the square loss term. In practice we select $\lambda$ from $10^{-5}$ to $10^{0}$ because this is below the variance of all datasets in our paper. Searching over different sparsity penalties ensures that our model search is not sacrificing quality of model fit in favor of model sparsity or vice versa.
    
    Finally, we fix a list of degree upper bounds $(D_1, \hdots, D_M)$ for all HDI models. For each choice of $\lambda$, we search over all degree combinations $(d_1, \hdots, d_m)$ where $1 \leq d_k \leq D_k$. 
    Since all hidden variables are indistinguishable, we only need to search over those degree combinations with $d_{m+1} \leq \hdots \leq d_M$.
    Because our focus in this paper is on stable oscillatory and chaotic systems, we do not allow $d_k = 0$ for any $1 \leq k \leq M$ as this would cause the coordinate $x_k$ to drift linearly with time.
    
    \item All initialized HDI models are dense which means they contain all polynomial terms up to degree $d_k$ in the $k$th equation for $1 \leq k \leq M$. Each randomly initialized model is parameterized by the stacked vector
    \begin{equation}
    \mathbf{p} = (\{b_k\}_{k=1}^M, \{\mathbf{c}_k\}_{k=1}^M, \{\tau_k\}_{k=1}^M)
    \end{equation}
    of its initial conditions, coefficients, and time scales and is trained for one round of optimization to minimize the objective
    \begin{equation}\label{eq:obj}
        L(\mathbf{p}) = \text{MSE}(\mathbf{p}) + \text{Reg}(\mathbf{p})
    \end{equation}
    which is a sum of the mean squared error
    \begin{equation}
        \text{MSE}(\mathbf{p}) = \frac{1}{n}\sum_{i=1}^n\sum_{j=1}^m(x_j(t_i, \mathbf{p}) - y_{ij})^2
    \end{equation}
    and a regularization term. The regularization term in the objective function is given by
    \begin{equation}
        \text{Reg}(\mathbf{p}) = \lambda\sum_{k=1}^M\sum_{|\bm\alpha| \leq d_k}\sqrt{1 + |\bm\alpha|}|c_{\bm\alpha}^k| + \gamma\sum_{k=1}^M (\|\mathbf{c}_k\|^2 - 1)^2
    \end{equation}
    where the first term penalizes the sparsity of the learned model, since polynomial terms $x_1^{\alpha_1}...x_M^{\alpha_M}$ in the model with a higher total degree $|\bm\alpha| = \alpha_1 + \hdots + \alpha_M$ are more actively down-weighted by the factor $\sqrt{1 + |\bm\alpha|}$. As stated previously, the sparsity penalty $\lambda$ is chosen between $10^{-5}$ and $10^0$. The second part of the regularization function above enforces that the coefficients $\mathbf{c}_k$ of the $k$th equation have a unit norm, which allows us to avoid scaling ambiguities between the timescales $\tau_k$ and model coefficients $\mathbf{c}_k$. The unit norm penalty on the coefficients is always set to the large value of $\gamma = 5 \times 10^{4}$, forcing the unit norm constraints to be satisfied almost exactly.
    
    Each model is optimized for 100,000 AdaBelief~\cite{zhuang2020adabelief} iterations followed by 50,000 iterations of BFGS~\cite{fletcher2013practical} (e.g. large enough for our optimization to converge). In general, our method is not overly sensitive to how these values are set.
    
    \item \rv{After one round of optimization, we remove in an automated fashion all models that do not pass the following dynamical criteria}.
    Namely, for oscillatory data such as the FHN model, neuron dynamics, and the BZ reaction, we only keep models whose fits to the data are periodic with amplitude and frequency commensurate with the recorded data. We check for periodicity by simulating a model trajectory up to time $T$, searching for the last time $t^* < T$ that the trajectory intersected its terminal point at time $T$, and then computing a sliding window distance to check whether this trajectory in $[t^*, T]$ was repeated in the two previous windows $[t^* - (r+1)(T-t^*), t^* - r(T-t^*)]$ for $r = 1, 2$.
    On the other hand, if the dynamics we aim to learn is chaotic instead of periodic such as the Lorenz attractor, we remove all models with periodic fits since periodicity is a very unlikely behavior for chaotic systems.
    In general, we also remove any models whose fits to the data converge to fixed points or diverge to infinity which is easily checked by simulating these models for a sufficiently long time window. We refer the reader to Fig.~\ref{fig:fhn_fit_types} which shows an example of a good HDI model fit to three oscillations of the $v$-coordinate of FHN as well as several models with unacceptable fits (convergent, divergent, incorrect period/amplitude) which we remove from our model search.
    
    \item For all remaining models, their relative errors (RE) on the \rv{training} data $\{(t_i, \mathbf{y}_i)\}_{i=1}^n$ of the observed variables $\mathbf{x}(t) = \{x_k(t)\}_{k=1}^m$ are computed using the formula
    \begin{equation}
        \text{RE} = \sqrt{\frac{1}{m}\sum_{k=1}^m\frac{\sum_{i=1}^n \Big(x_k(t_i) - y_{ik}\Big)^2}{\sum_{i=1}^n \Big(y_{ik} - \frac{1}{n}\sum_{j=1}^n y_{jk}\Big)^2}} \, .
    \end{equation}
    
    Now we need to decide which models have a sufficiently good fit by keeping only those whose train RE is below a certain threshold $\theta_{RE}$. To do this, we fit a Gaussian kernel density estimator to the REs of all models where the kernel bandwidth is set through least squares cross validation.
    We then use peak prominence to choose the first minima in the train loss density with prominence above 0.01 (i.e. a robust heuristic for detecting the first significant drop) and set this value as our RE cutoff $\theta_\text{RE}$. In Fig.~\ref{fig:train_histogram_dists}(a) we show the histogram of HDI model REs on \rv{training} data generated from the FHN system with 30\% additive noise. The kernel density estimator fit to the histogram is shown in dark blue and the black vertical dashed line denotes the cutoff threshold $\theta_\text{RE}$ determined through peak prominence.
\end{enumerate}

\begin{figure}
    \centering
    \includegraphics[width=\textwidth]{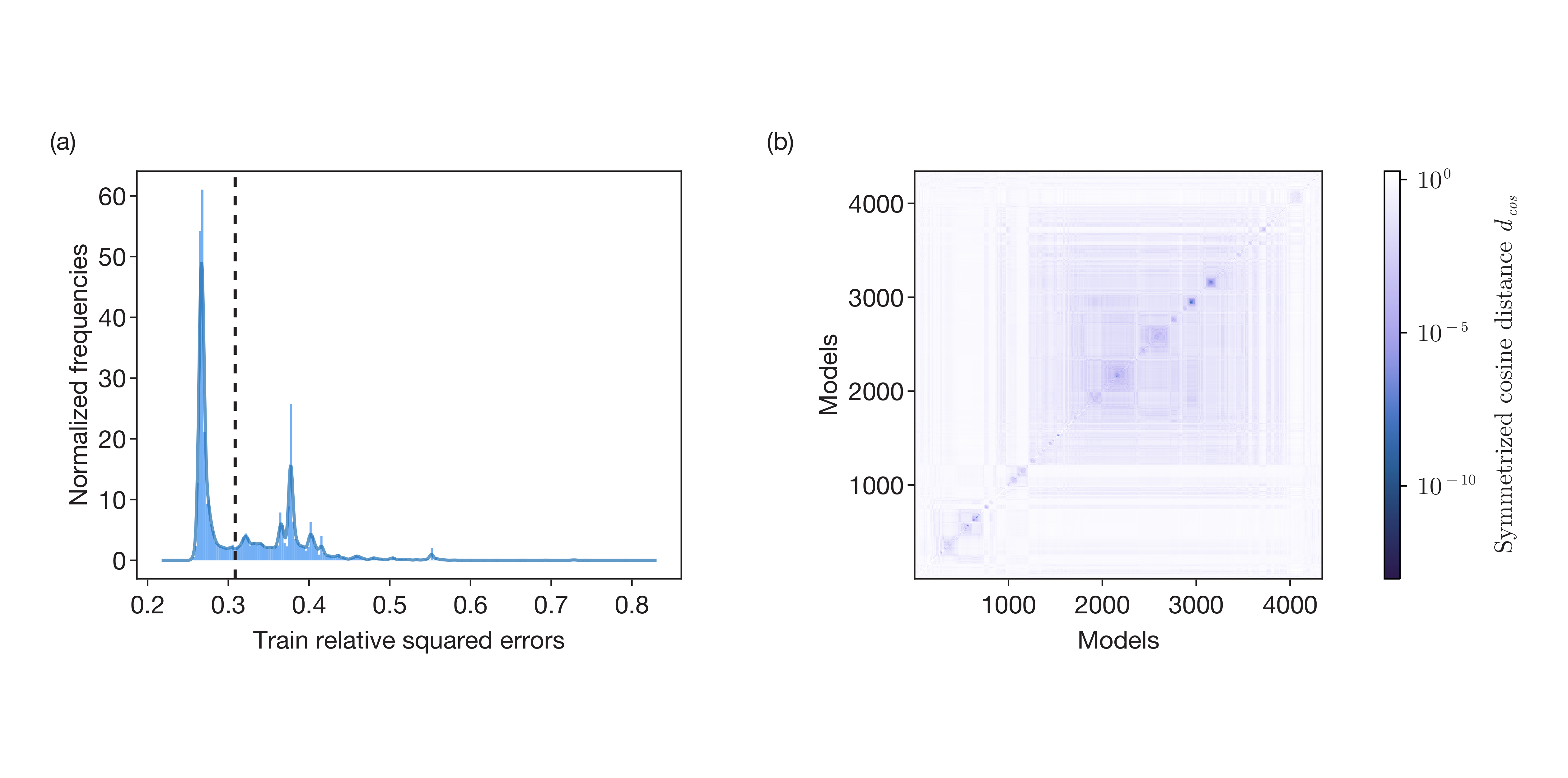}
    \caption{(a) Train RE histogram for FHN system with 30\% noise and (b) symmetrized cosine distance between all remaining models}
    \label{fig:train_histogram_dists}
\end{figure}

\par\textbf{Model Distance Matrix}
\begin{enumerate}
    \setcounter{enumi}{5}
    \item For all remaining models with RE below the threshold $\theta_{RE}$, we rescale their coefficients $c_{\bm\alpha}^k$ and time scales $\tau_k$ such that the standard deviations of the trajectories all hidden variables $h_{m+1}, \hdots, h_M$ are equal to the standard deviation of the trajectory of the first observed variable $x_1$. This normalization ensures that the standard deviations of the hidden variables across all learned models are on the same scale and hence, the coefficients of these models can be compared to each other.
    For all models whose equation degree $d_k$ is strictly less than the maximal degree $D_k$, we pad their coefficients with zeros to ensure that all models have equations of maximal degree $(D_1, \hdots, D_k)$.
    This padding ensures that the coefficient vectors of all models are of the same length. After normalization and padding are performed, we extract from each model the stacked coefficient vector $\overline{\mathbf{c}} = \{\tau_k\mathbf{c}^k\}_{k=1}^M$ with timescales included. Now each model is fully described by its stacked vector of coefficients $\overline{\mathbf{c}}$ and this representation can be used to compare it to other models.
    
    \item To compare the learned hidden variable models, we  account for certain transformations of the hidden variables, specifically sign flips and permutations, that result in equivalent models.
    Starting with a model with stacked coefficient vector $\overline{\mathbf{c}}$ which corresponds to the variables $(x_1, \hdots, x_m, h_{m+1}, \hdots, h_k, \hdots, h_M)$ whose degree combination is $(d_1, \hdots, d_M)$.
    For any $k_1, \hdots, k_p \in \{m+1, \hdots, M\}$ with $p \leq M-m$, we denote $\sigma_{k_1, \hdots, k_p}(\overline{\mathbf{c}})$ as the stacked coefficient vector of the equivalent model with variables $(x_1, \hdots, x_m, h_{m+1}, \hdots, -h_{k_1}, \hdots, -h_{k_2}, \hdots, -h_{k_p}, \hdots, h_M)$ where all hidden variables at positions $k_1, \hdots, k_p$ have their signs flipped. Note that the sign flip positions $k_1, \hdots, k_p$ do not have to be contiguous. Also, for any permutation $\pi$ of the hidden variables $(h_{m+1}, \hdots, h_M)$ which satisfies $\pi(d_{m+1}, \hdots, d_M) \leq (D_{m+1}, \hdots, D_M)$,
    we denote $\pi(\overline{\mathbf{c}})$ as the stacked coefficient vector of the equivalent model with variables $(x_1, \hdots, x_m, \pi(h_{m+1}, \hdots, h_M))$.
    
    \item Finally, we compute a pairwise distance matrix using the cosine distance between every pair of models (stacked coefficient vectors) $\overline{\mathbf{c}}, \overline{\mathbf{c}}'$ modulo these sign flip and permutation symmetries. The cosine distance is upper bounded by 2 which makes it more amenable to thresholding and model clustering as compared to Euclidean distance. Specifically, we use the cosine distance minimized over all possible sign flips and permutations of the hidden variables given by
    \begin{equation}
        d_{\text{cos}}(\overline{\mathbf{c}}, \overline{\mathbf{c}}')
        = \min \Big\{
        1 - \frac{
            \inprod{
                \overline{\mathbf{c}}, \pi(\sigma_S(\overline{\mathbf{c}}'))
            }
        }{
            \| \overline{\mathbf{\mathbf{c}}} \|
            \| \pi(\sigma_S( 
                \overline{\mathbf{c}}' 
            ) ) \| 
        }
        \Big| \ S \in 2^{\{m+1, \hdots, M\}}, \quad \pi: \pi(d_{m+1}, \hdots, d_M) \leq (D_{m+1}, \hdots, D_M)\Big\}
    \end{equation}
    where $2^{\{m+1, \hdots, M\}}$ denotes the powerset (the set of all subsets) of $\{m+1, \hdots, M\}$.
    
    In Fig.~\ref{fig:train_histogram_dists}(b), for the example of FHN with 30\% noise, we plot the $d_\text{cos}$ distance between all $N=4,432$ models whose train REs are below the cutoff $\theta_\text{RE}$. These $N=4,432$ models correspond to all the models in the train RE histogram to the left of the vertical dashed line in Fig.~\ref{fig:train_histogram_dists}(a).
\end{enumerate}

\begin{figure*}
    \centering
    \includegraphics[width=\textwidth]{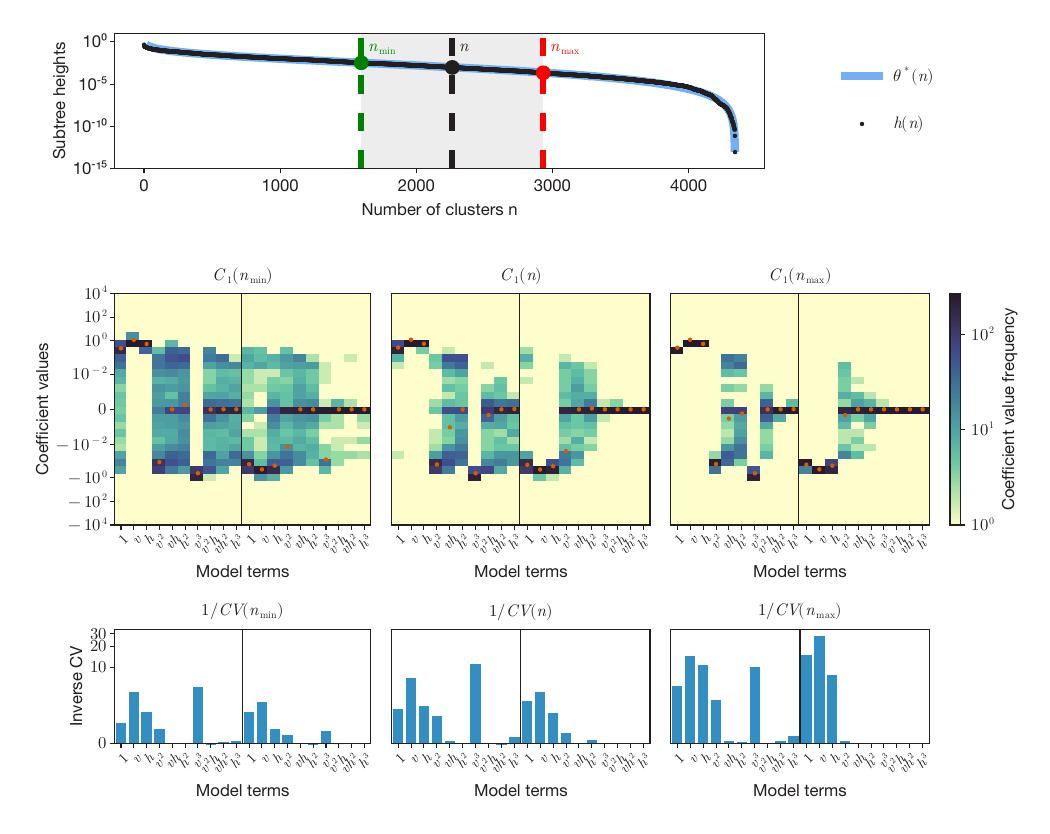}
    \caption{Monotonic curve of subtree heights $h(n)$ from hierarchical clustering of models learned on FHN data with 30\% noise (top). The resulting dominant clusters $C_1(n_{\text{min}}), C_1(n),$ and $C_1(n_{\text{max}})$ shown for three different numbers of clusters $n_{\text{min}} \leq n \leq n_{\text{max}}$ (middle). At each cutoff $n$ we take the top cluster $C_1(n)$, and for each model term $1, v, h, ...$ we collect the coefficients for that term across all models in the cluster and compute the coefficient of variation by formula~\eqref{eq:cv} (bottom). As expected, we find that the true FHN terms ($1, v, h, v^3$ and $1, v, h$) have the largest coefficients with the smallest spread and hence have the largest inverse coefficient of variation score.}
    \label{fig:hierarchical_clusters}
\end{figure*}

\par\textbf{Model Clustering}
\begin{enumerate}
    \setcounter{enumi}{8}
    \item Using this distance matrix between all models, we perform single-linkage clustering and obtain a dendogram which specifies how models are sequentially grouped together into clusters.
    
    \item We now need to determine where to cut the dendogram tree generated by hierarchical clustering to split our models into separate clusters. Choosing a cutoff too high or too low in the dendogram results in clusters which are overestimated (too many models) or underestimated (too few models) respectively. Here we describe a robust procedure for choosing an interval of cutoffs $[h_\text{min}, h_\text{max}]$ that give reliable model clusters.
    
    First, from the hierarchical clustering output we extract the non-decreasing sequence of subtree heights $h(n)$ for $1 \leq n \leq N$ where $N=4,432$ is the total number of models in our model set from the initial optimization step. By definition, cutting the dendogram formed by hierarchical clustering at height $h(n)$ should return $n$ separate model clusters. The function $\log_{10}h(n)$ has a sigmoidal shape which is well approximated by the generalized logistic function or Richard's curve
    \begin{equation}
        \theta(n) = \Big(1 + a\exp\Big(-c(n/N-b)\Big)\Big)^{-d}.
    \end{equation}
    Given a decreasing sequence of log subtree heights $\log_{10}h(n)$ from our hierarchical clustering, we fit the parameters $a, b, c, d$ of our Richard's curve above by initializing them randomly in $[0, 1]$ and optimizing through gradient descent to minimize the least squares objective $\sum_{n=1}^N (\theta(n) - \log_{10}h(n))^2$. Obtaining the least squares fit $\theta^*(n)$, we then compute the unique inflection point of this analytic curve where the second derivative is zero which we denote $h_{\text{min}}$ and the unique elbow point $h_{\text{max}} > h_{\text{min}}$ where the third derivative is zero.
    This procedure allows us to select an intermediate region of subtree heights $[h_{\text{min}}, h_{\text{max}}]$ where the model groups determined by hierarchical clustering are reliable. Since $h(n)$ is an invertible decreasing function, we can also compute $n_{\text{min}} = h^{-1}(h_{\text{max}})$ and $n_{\text{max}} = h^{-1}(h_{\text{min}})$ which gives us the smallest and largest number of clusters respectively which we could reliably split our $N$ models into.

    In the top of Fig.~\ref{fig:hierarchical_clusters} we plot in black the curve of subtree heights $h(n)$ from a hierarchical clustering of the $N = 4,432$ models learned from a single simulation of the FHN $v$-coordinate with $30\%$ additive noise. We also show the least-squared fit Richard's curve $\theta^*(n)$ in blue which allows us to determine the inflection points $h_{\text{max}}, h_{\text{min}}$ of $h(n)$ which occur at the cluster numbers $n_{\text{min}}, n_{\text{max}}$ shown in green and red.
    
    \item Our goal now is to determine the largest model cluster of all $N$ models in our set. However, at each cutoff threshold $h_{\text{min}} \leq h \leq h_{\text{max}}$ the clusters returned by hierarchical clustering can change their size and contents. Here we describe how to choose the model cluster that is largest \textit{on average} for all cutoffs in the range $[h_{\text{min}}, h_{\text{max}}]$. First, let $C_p(n) \subseteq \{1, \hdots, N\}$ denote the unique cluster of models which contains model $p \in \{1, \hdots, N\}$ at cutoff $h(n)$. Fixing a root model $p \in \{1, \hdots, N\}$, for all $n_{\text{min}} \leq n \leq n_{\text{max}}$ we count the number of times the cluster $C_p(n)$ is the largest compared to all other clusters $C_q(n))$ for $q \neq p$ at that same dendogram height (cutoff level). Then we choose the first model $p$, in numerical order, which has the highest count (falls into the top cluster most frequently). Without loss of generality assume that $p = 1$. Finally, this gives us a sequence of nested model clusters \begin{equation}
        C_1(n_\text{min}) \supseteq \hdots \supseteq C_1(n) \supseteq \hdots \supseteq C_1(n_\text{max}).
    \end{equation}
    These nested clusters effectively trace a path up the dendogram tree which is initialized at root $p=1$ at the bottom of the tree. On average, at any cutoff $n_{\text{min}} \leq n \leq n_{\text{max}}$, the cluster $C_1(k)$ contains the largest number of models (i.e. is the dominant cluster).
    
    In the middle of Fig.~\ref{fig:hierarchical_clusters}, we plot the dominant (largest) cluster $C_1(n)$ given that we perform hierarchical clustering with $n$ total clusters. The heatmap for each cluster shows for each polynomial model term ($1, v, h$, ...) the distribution of its coefficient values for all the models in that cluster. Note that as the number of clusters $n$ in our hierarchical clustering increases, the dominant cluster $C_1(n)$ (Fig.~\ref{fig:hierarchical_clusters} middle) becomes smaller and its model coefficients concentrate more tightly around the true polynomial terms of the FHN oscillator as expected.
\end{enumerate}

\par
\begin{figure*}
    \centering
    \includegraphics{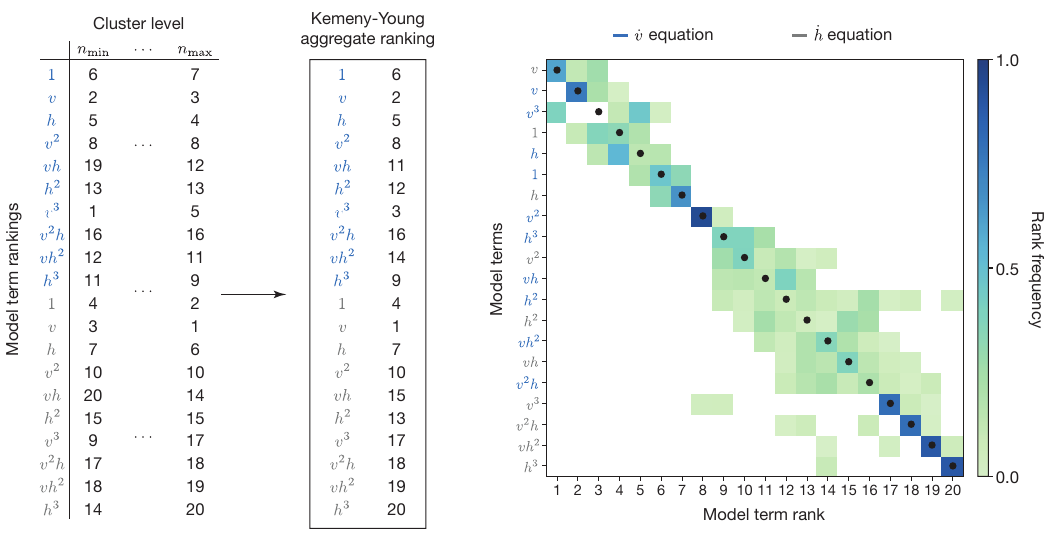}
    \caption{Here we show how model terms are ranked for two-variable HDI models fit to simulated FHN data with 30\% additive noise. In the leftmost table we show the list of rankings $\{(r_1^n, \hdots, r_T^n)\}$ in each column for every cluster level $n = n_{\text{min}}, \hdots, n_{\text{max}}$. This then is aggregated into a final model term ranking using the Kemeny-Young algorithm shown in the middle display. In the rightmost plot we visualize the rank order for each term \rv{written on the left, where a blue color term indicates it} belongs to the equation for $\dot{v}$ and a \rv{gray color term} indicates \rv{it} belongs to the equation for $\dot{h}$. Each row of the heatmap indicates the frequency with which a given term is assigned a certain rank across all cluster levels $n = n_{\text{min}}, \hdots, n_{\text{max}}$.}
    \label{fig:term_rankings}
\end{figure*}

\textbf{Model Term Ranking}
\begin{enumerate}
    \setcounter{enumi}{10}
    \item Now that we have obtained the nested sequence of dominant clusters from our hierarchical clustering, we are able to study the models appearing in the clusters $C_1(n)$ at every level $n_{\text{min}} \leq n \leq n_{\text{max}}$ and sort the terms in these models from most to least important. At a high level, the importance of a model coefficient is measured by its magnitude and variability across all models in a given cluster. Terms that are consistently equal to the same large value across all models will be ranked higher (more important) whereas terms which vary greatly or are close to zero will be ranked lower (less important).
    
    Now we describe exactly how this model term ranking is performed. At a given level $n$, we take the cluster $C_1(n)$ and compute the stacked coefficient vectors $\overline{\mathbf{c}} = \{\tau_k\mathbf{c}^k\}_{k=1}^M$ for every model in this cluster. Assume that in this cluster we have $R$ models denoted by their stacked coefficient vectors $\overline{\mathbf{c}}_1, \hdots, \overline{\mathbf{c}}_R$. We would like to understand which entries (model terms) across all $r$ vectors are large in magnitude and have low variability. In order to do this, we first need to align the model vectors by removing all possible sign flips and permutations due to the symmetries of the hidden variables. Without loss of generality, we fix the first model $\overline{\mathbf{c}}_1$ as reference and align all other models for $2 \leq r \leq R$ by setting
    \begin{equation}
        \overline{\mathbf{c}}_r \to \overline{\mathbf{c}}_r' = \pi(\sigma_S(\overline{\mathbf{c}}_r))
    \end{equation}
    where the hidden variable sign flips $S$ and permutation $\pi$ are chosen to minimize the cosine distance to the first model
    \begin{equation}
        S, \pi \in \arg\min\Big\{1 - \frac{\inprod{\overline{\mathbf{c}}_1, \pi(\sigma_S(\overline{\mathbf{c}}_r))}}{\|\overline{\mathbf{\mathbf{c}}}_1\|\|\pi(\sigma_S(\overline{\mathbf{c}}_r))\|} \ \Big| \ S \in 2^{\{m+1, \hdots, M\}}, \quad \pi: \pi(d_{m+1}, \hdots, d_M) \leq (D_{m+1}, \hdots, D_M)\Big\} \, .
    \end{equation}
    Now assume that all models $\overline{\mathbf{c}}_1, \hdots, \overline{\mathbf{c}}_R$ are aligned as described above. The $i$th entry $\overline{\mathbf{c}}_{ir}$ in the stacked coefficient vector for models $r = 1, \hdots, R$ corresponds to a polynomial term in one of the $M$ equations of our ODE system. To test whether this term is important for the dynamics learned by these $R$ models, we compute its coefficient of variation given by
    \begin{equation}\label{eq:cv}
        \text{CV}_i = \Big(1 + \frac{1}{\sqrt{2(R-R_0^i)}}\Big)\frac{\text{interquartile range}(\{\overline{\mathbf{c}}_{ir}\}_{r=1}^R)}{\text{median}(\{\overline{\mathbf{c}}_{ir}\}_{r=1}^R)} \, .
    \end{equation}
    The interquartile range of a set of numbers is defined as the difference between the 75th and 25th percentiles. The median is defined as the 50th percentile. Here $R_0^i$ is the number of times the coefficient $\overline{\mathbf{c}}_{ir}$ for model $r$ in the $i$th position is exactly equal to zero for $r = 1, \hdots, R$. The discount factor of $1 +  \frac{1}{\sqrt{2(R-R_0^i)}}$ increases the coefficient of variation for terms which are frequently set to zero and is only important when models with different degree combinations (sparsity patterns) are grouped into one cluster.
    Intuitively, if a term $i$ is consistent and tightly clustered around a large mean value for all models $\overline{\mathbf{c}}_{i1}, \hdots, \overline{\mathbf{c}}_{iR}$, then it receives a small coefficient of variation. Otherwise, if a term $i$ has large spread or is close to zero across all $R$ models, then it receives a large coefficient of variation.
    
    \item Finally, for the top cluster $C_1(n)$ at every level $n_{\text{min}} \leq n \leq n_{\text{max}}$, using this definition we can compute the coefficients of variation for every model term denoted by $\text{CV}_1(n), \text{CV}_2(n), \hdots, \text{CV}_T(n)$ where $T$ denotes the total number of terms in our ODE system. In the bottom of Fig.~\ref{fig:hierarchical_clusters}, for models learned on simulated FHN data with 30\% noise, we show the coefficients of variation for all model terms at three different clustering levels $n_{\text{min}} \leq n \leq n_{\text{max}}$. We note that the correct FHN model terms appear with the lowest coefficients of variation (highest inverse values). As the cluster level $n$ increases, the coefficients of variation decrease the fastest around the true model terms as expected.
    
    From this we can obtain a ranking $r_1^n, \hdots, r_T^n$ of the $T$ terms in our model by sorting the coefficients of variation from smallest to largest (breaking ties based on numeric order). Here $r_1^n, \hdots, r_T^n$ is simply a permutation of the set $\{1, \hdots, T\}$. To robustly determine the ranking of model terms from most to least important, we create a consensus from the individual rankings obtained at each cluster level $\{(r_1^n, \hdots, r_T^n)\}_{n = n_{\text{min}}}^{n_{\text{max}}}$. To aggregate these rankings into one, we use a  traditional method for aggregating votes or ballots known as the Kemeny-Young or VoteFair popularity ranking algorithm~\cite{kemeny1959mathematics, young1995optimal}. This algorithm finds the \textit{average} ranking that minimizes the sum of Kendall tau distances to the list of individual rankings at each cluster level $n = n_{\text{min}}, \hdots, n_{\text{max}}$. Using this method, we obtain the final aggregated ranking $r_1, \hdots, r_T$ of all $T$ terms in our ODE model.

    In Fig.~\ref{fig:term_rankings} for the example of FHN with 30\% noise we display the list of rankings $\{(r_1^n, \hdots, r_T^n)\}$ at every cluster level $n = n_{\text{min}}, \hdots, n_{\text{max}}$ and give an example of how the  Kemeny-Young algorithm aggregates these results into a final model term ranking for FHN.
\end{enumerate}

\begin{figure*}
    \centering\includegraphics[width=\textwidth]{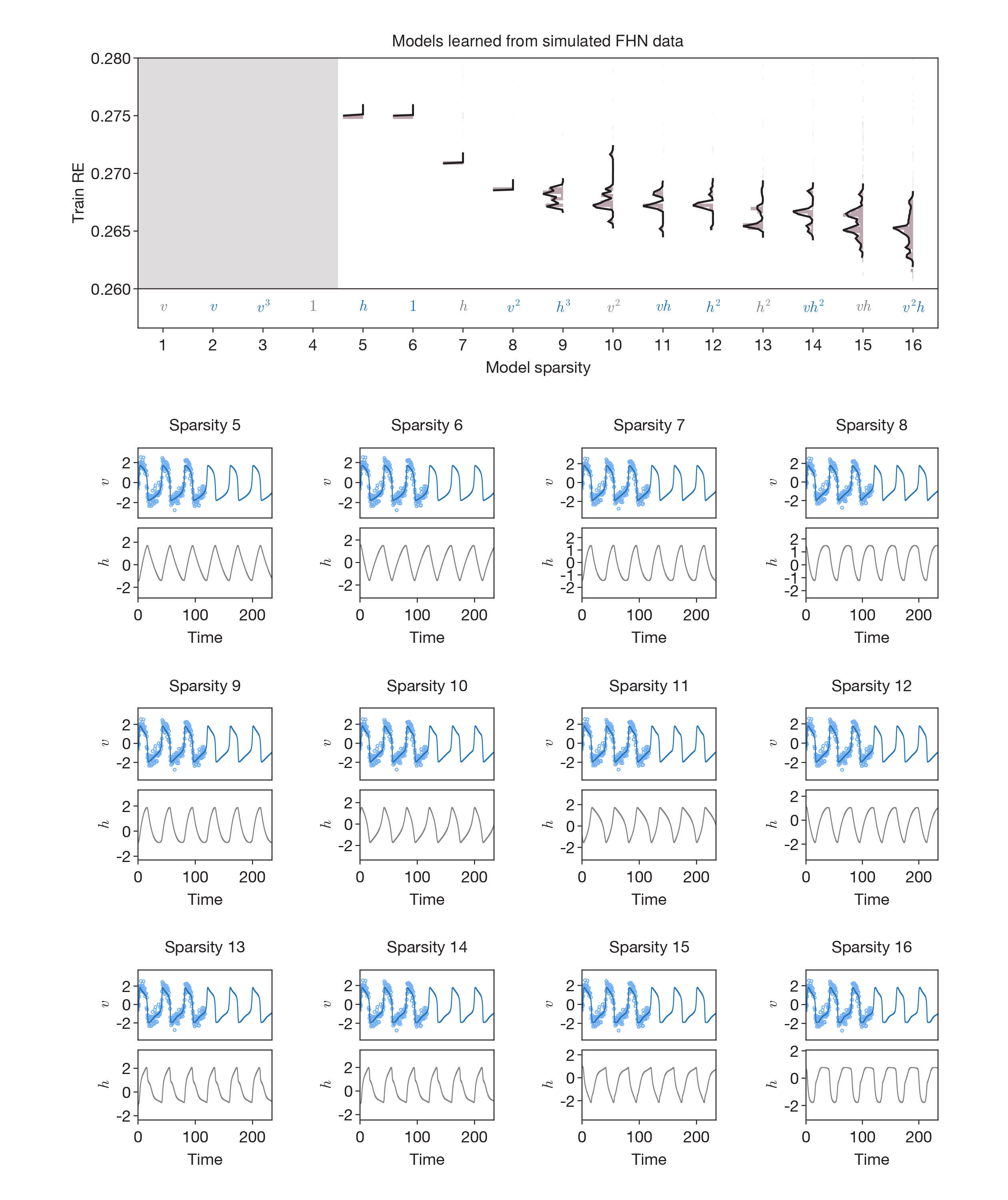}
    \caption{Given training data consisting of three periods of oscillation for the FHN $v$-coordinate corrupted with 30\% noise, our HDI model search repeatedly fits two-variable cubic ODE models and learns a ranking of model terms shown in the top panel. Following our model sparsification procedure, for every sparsity $s$, the top $s$ terms in our ranking are combined into a new model and refit to the \rv{training} data from 2000 random parameter initializations. At each sparsity $s$, the RE on the \rv{training} data across all learned models is plotted as a histogram (after all nonperiodic models are removed). We observe that \rv{the first drop and largest gap} in the RE is observed at sparsity 7, which contains all the terms of the true FHN model ($1, v, h, v^3$ and $1, v, h$).}
    \label{fig:fhn_sparsity}
\end{figure*}

\par\textbf{Model Sparsification}
\begin{enumerate}
\setcounter{enumi}{12}
    \item Now that we have obtained a final ranking of model terms $r_1, \hdots, r_T$ we proceed to the final step of finding a sparse model that fits our observed data. To do this, for every sparsity $s = 1, \hdots, T$ we take the top $s$ terms in our ranking (i.e. the polynomial terms at indices $i_1, \hdots, i_s$ for which $r_{i_1}, \hdots, r_{i_s} \in \{1, \hdots, s\}$). We then refit this sparse $s$ term model from 2000 different random initializations of its parameters (coefficients, initial conditions, and timescales). From these optimizations, only those models which satisfy the criteria described in Fig.~\ref{fig:fhn_fit_types} are kept (see point 4 of section ``Initial Optimization" above). The RE train loss of all kept models at each sparsity level $s$ can then be aggregated into a histogram. The user of our framework can now study how the train loss statistics of models at each sparsity level (e.g. mean RE, 10th percentile RE, etc.) decrease as more terms are added to the model (as $s$ increases). Because real-world dynamical systems are rarely of polynomial form, the choice of how many polynomial terms $s$ to keep in the final learned model can and should be based on the scientific judgement of the user and their preference for model complexity vs. goodness of fit.

    In Fig.~\ref{fig:fhn_sparsity} we show several examples of sparse HDI models fit to simulated data of 3 periods of the FHN $v$-coordinate with 30\% additive Gaussian noise. We plot how the histogram of model train losses change as more terms are added into the model based on their ranked order from Fig.~\ref{fig:term_rankings}. The true FHN system in this model term ranking is recovered at sparsity 7. The sparsest models that fit the observed oscillations in $v$ have 5 and 6 terms and are equivalent to a constantly forced Van der Pol oscillator which, in the right parameter regime, can have a stable limit cycle and can even be excitable (long excursions that tend to a fixed point)~\cite{strogatz2018nonlinear}. Hence, these 5 and 6 term models are also valid models for neuron dynamics that share the same properties as the FHN model.

    \item Finally, using this same procedure for HDI optimization, the final sparse model chosen by the user can now be refit to new test recordings such as data collected from different systems or observations of the same system under different environmental conditions. The user can now study bifurcations of their learned model and classify different datasets based on the learned ODE coefficients that describe them.
\end{enumerate}

\begin{figure*}
    \centering
    \includegraphics{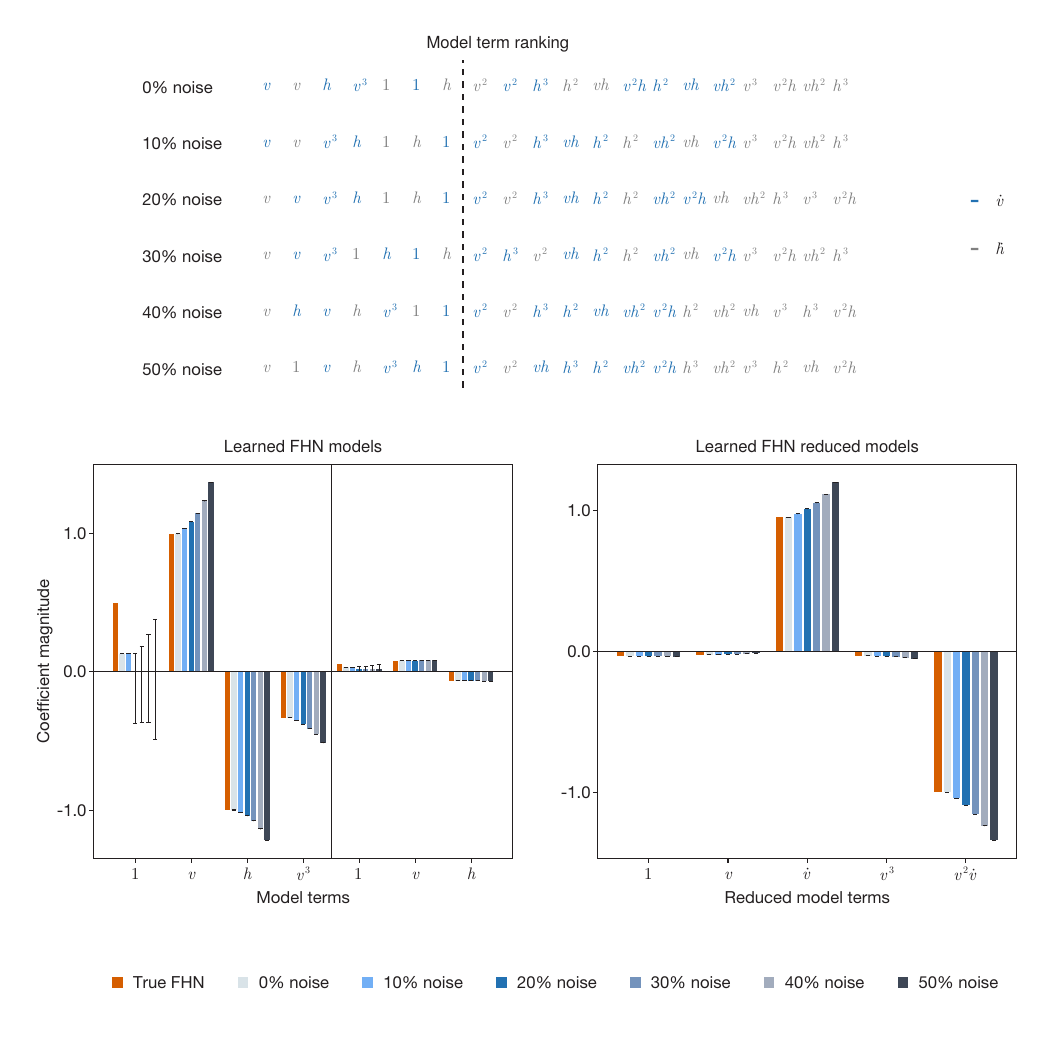}
    \caption{From just three oscillations of the $v$-coordinate of FHN, even with up to 50\% noise corruption we find that all seven terms of the true FHN model are correctly ranked as the top seven terms by the ``Model Term Ranking" step of the  HDI model search procedure (top panel). Keeping the top seven terms in our ranking (which are the true FHN terms), we run 2000 optimizations where we refit this seven term sparse model from random initialization to the same noisy simulated data of the FHN $v$-coordinate (see ``Model Sparsification" step). The bottom left panel displays the coefficient values of these seven term models when they are refit to this data. Naturally, since the hidden variable $h$ is unobserved, the coefficients of the bias terms (e.g. terms $1$) in both $\dot{v}$ and $\dot{h}$ equations do not agree with the true simulated FHN coefficients even with 0\% added noise. Transforming the coefficients of every seven term model into their reduced form (see section ``FHN model") we find that all learned models do in fact agree with the reduced coefficients (e.g. have learned the correct equations) of the ground-truth FHN system and vary smoothly with the increase of noise in the data.}
    \label{fig:fhn_noise_robustness}
\end{figure*}

Our model search pipeline described above was shown on the example of the simulated FHN system with 30\% additive noise in its first coordinate. In Fig.~\ref{fig:fhn_noise_robustness} we show how the model inferred by our search procedure varies when the amount of additive noise is taken from 0\% to 50\%. We see that our framework is robust to high levels of noise in the data as it learns the correct terms in the $(v, h)$ FHN system as well as the correct higher order dynamics in $(v, \dot{v})$.

\begin{figure*}
    \centering
    \includegraphics{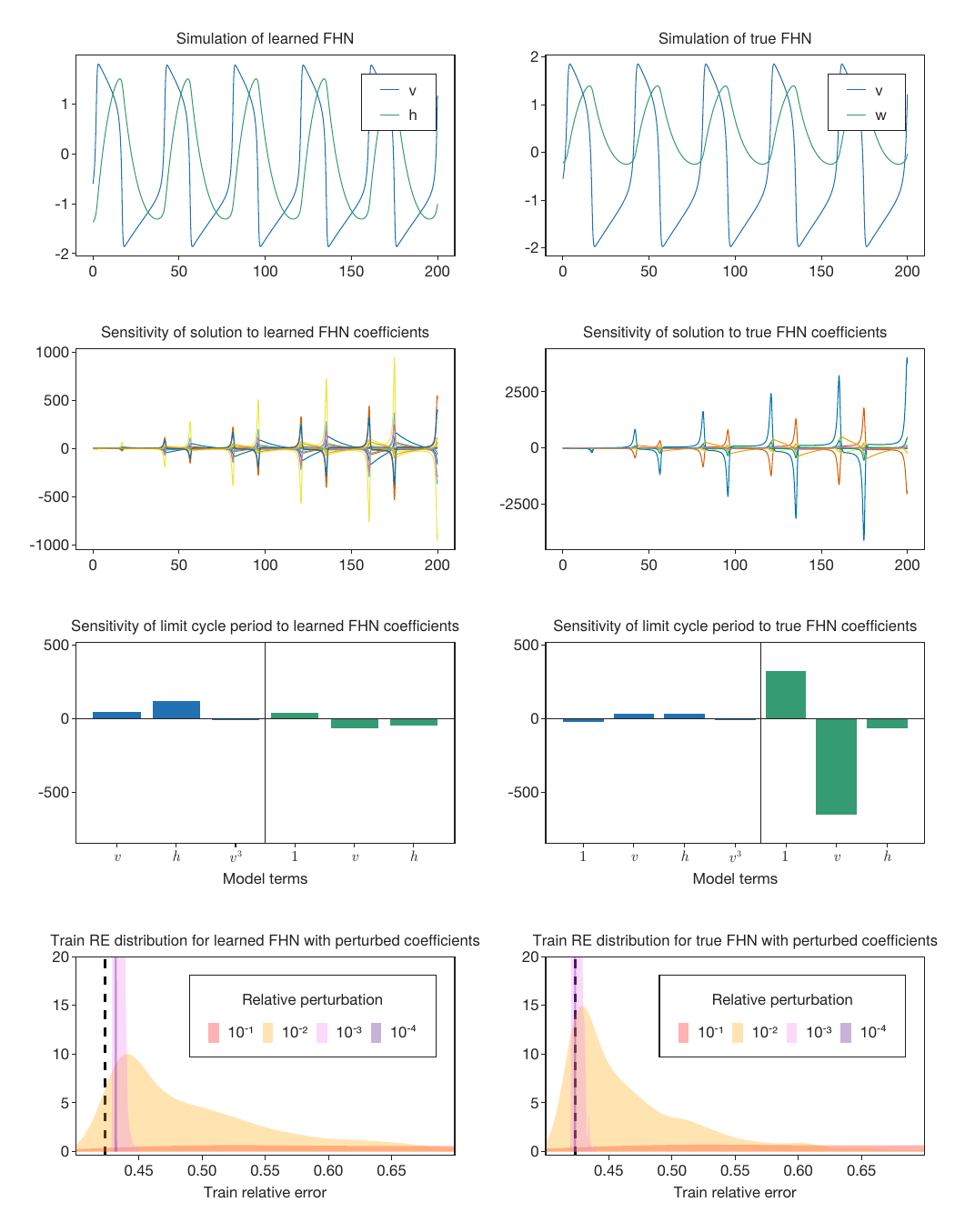}
    \caption{\rv{Sensitivity of learned FHN model with 50\% noise from main text Fig.~1 compared to true FHN model. Both models have timescales set to one (i.e. multiplied into their coefficients). Top row shows trajectories of both variables in learned and true model. Second row shows sensitivities of these trajectories with respect to model coefficients and initial conditions. Third row computes the sensitivities of limit cycle period in both models with respect to their coefficients as described in~\cite{larter1984sensitivity}. Fourth row plots the distribution of relative errors on training data when coefficients of both models are perturbed by $\mathbf{p} \mapsto \mathbf{p} \odot (1 + \sigma \cdot \boldsymbol{\eta})$ for a standard normal vector $\boldsymbol{\eta}$ where $\sigma = 10^{-1}, 10^{-2}, 10^{-3}, 10^{-4}$.}}
    \label{fig:fhn_sensitivity}
\end{figure*}

\clearpage

\section{Neuron models from squid axon data}
\begin{figure*}
    \centering
    \includegraphics[width=\textwidth]{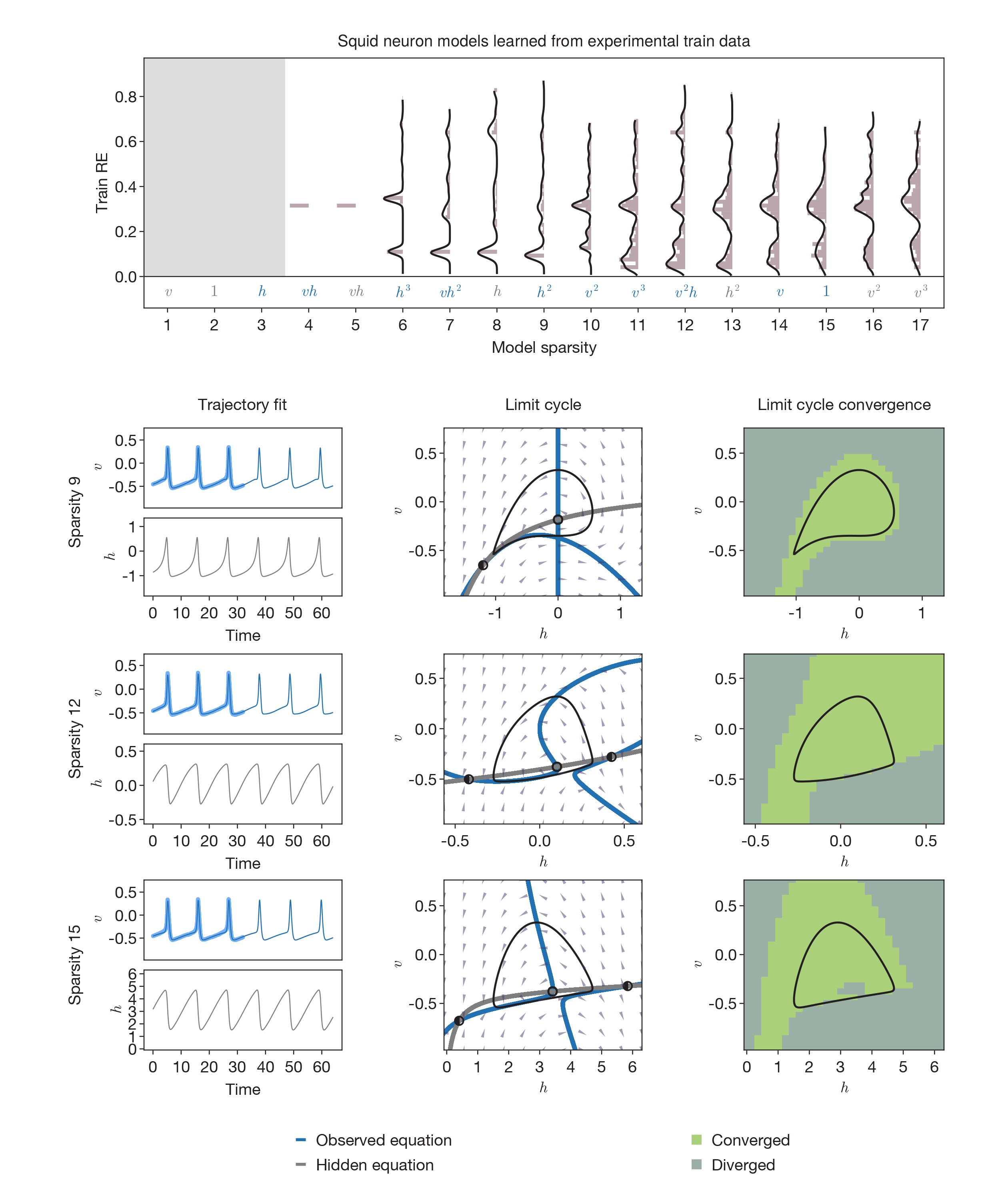}
    \caption{From three periods of a spike train experimentally recorded from a squid axon, our method repeatedly fits two-variable ODE models with all cubic terms in $(v, h)$ to this \rv{training} data and learns a ranking of model terms shown in the top panel. At each sparsity level $s$, the top $s$ terms are combined into a new model and its coefficients are refit to the \rv{training} data from 2000 random parameter initializations. The losses on the \rv{training} data for all refit models of a given sparsity are plotted as a histogram in the top panel. The next three rows display the learned models with 9, 12, and 15 terms respectively. The first column displays the model fit to the data, the second column shows the \rv{stable} limit cycle, nullclines, fixed points, and vector variables of their phase plane dynamics, and the third columns marks in bright green the region of initial conditions in the phase plane that converge onto the limit cycle of each model.}
    \label{fig:squid_neuron_sparsity}
\end{figure*}

\begin{figure*}
    \centering
    \includegraphics[width=\textwidth]{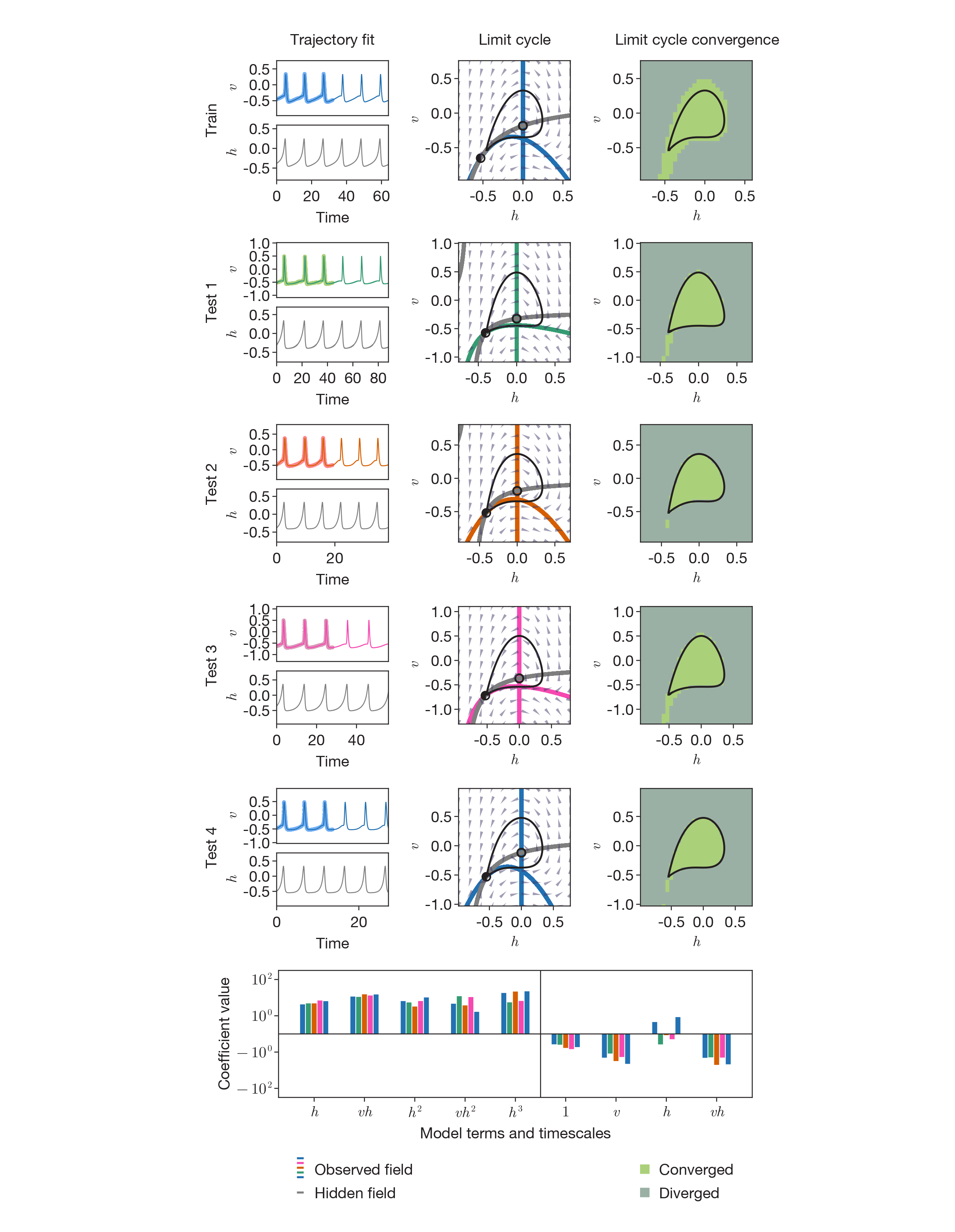}
    \caption{HDI model with nine terms fit to train data squid axon recording as well as four additional test recordings from new squid axons (first column). Fit models display a \rv{stable limit cycle approaching a homoclinic orbit} with nullclines $\dot{c} = 0$ and $\dot{h} = 0$ shown in colored and gray lines respectively (center column). In the third column we show how all learned models have a large region of convergence where their dynamics tends to the stable limit cycle when started from a new initial condition (bright green squares). The coefficients of all models are consistent across all five squids (bottom plot).}
    \label{fig:squid_neuron_train_and_test}
\end{figure*}

Now that we have described in detail our HDI model sweep procedure and shown its application to the simulated neuron model of FHN, we apply our framework to real experimental measurements~\cite{paydarfar2006noisy, physionet2000} of giant axon membrane potentials of the North Atlantic longfin inshore squid (Loligo pealeii) stimulated by noisy input currents. Consistent with prior neuron models~\cite{fitzhugh1961impulses, morris1981voltage, hodgkin1952currents, gerstner2014neuronal}, we search over two-variable HDI models in $(v, h)$ with at most cubic polynomial nonlinearities in both coordinates. In Fig.~\ref{fig:squid_neuron_sparsity} we show the model term ranking discovered by our sweep as well as the HDI models that result when we keep the top 9, 12, and 15 polynomial terms respectively. The phase plane dynamics of all models are governed by a stable limit cycle approaching a homoclinic orbit. In Fig.~\ref{fig:squid_neuron_train_and_test} we show how our 9 term HDI model is able to fit new experimental recordings of axon membrane potentials from different squids with ODE coefficients that remain consistent across all examples.

\section{Belousov-Zhabotinsky reaction: Chemical Experiments and Models}

\begin{figure*}\centering\includegraphics[width=\textwidth,height=\textheight,keepaspectratio]{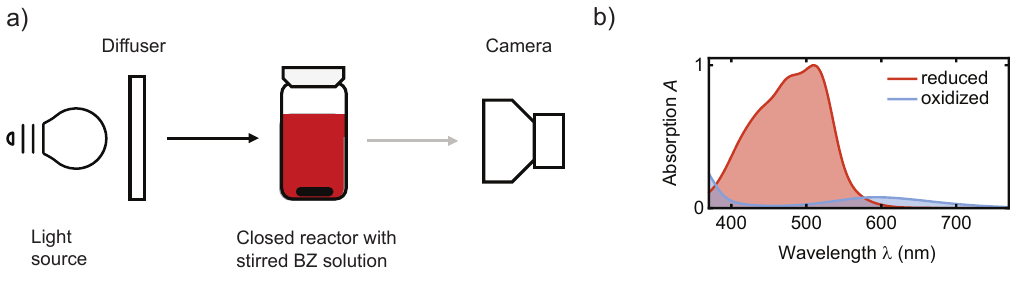}
    \caption{Optical measurement of periodic concentration changes in the oscillating  Belousov-Zhabotinsky reaction. a) The experimental setup consists of a spatially homogenized broadband light-source supplying the illumination that is absorbed by the reagents in the closed stirred chemical. The transmitted light is captured by a camera. b) During chemical oscillations the ferroin reagent cyclically changes its oxidation state affecting its corresponding absorption spectrum: reduced catalyst \ch{Fe^{2+}} (red) and oxidized catalyst \ch{Fe^{3+}} (blue). This allows for tracking the chemical oscillation state optically.}
    \label{fig:SI_bz_stability}
\end{figure*}

The Belousov-Zhabotinsky (BZ) reaction is the paradigmatic nonlinear chemical oscillator. It involves more than 30 chemical species and 40 elementary reactions. Over the course of the net reaction an organic substrate such as malonic acid (MA) is consumed:
\begin{align}
    \ch{3 MA + 2 BrO3- + 2 H^+ -> 2 BrMA + 3 CO2 + 4 H2O} \, .
\end{align}
However, intermediary species (\ch{Br-}, \ch{BrO3+}, \ch{[Fe(phen)3]^{3+}}) are periodically built up and expended, leading to a periodically changing consumption rate of the organic substrate. One of the intermediary species, the oxidized form of the catalyst ferroin \ch{[Fe(phen)3]^{3+}}, can be readily observed optically using spectrophotometry (Fig. \ref{fig:SI_bz_stability}a) to track the chemical oscillation state.

In our experiments we freshly prepare the BZ reaction in a \SI{50}{\milli\litre} vial from stock solutions of \SI{1}{\Molar} malonic acid, \SI{2.5}{\Molar} sulfuric acid, \SI{25}{\milli\Molar} ferroin, \SI{1}{\Molar} sodium bromate and \SI{1}{\Molar} sodium bromide (Millipore Sigma). 
Used concentrations are listed in the caption of Fig. 3 in the main text.
To keep the solution spatially homogeneous during the reaction we mix the solution with a magnetic stirrer at 1500 revolutions per minute (IKA Lab disc S41). 
For the spectrophotometry, we employ spatially uniform white background illumination provided by a light emitting diode (LED) lightbulb (800 lumens, \SI{10.5}{\watt}) and a light diffuser.
Light transmitted through the BZ solution vial is recorded with a complementary metal–oxide–semiconductor (CMOS) camera (Canon EOS 200D).
The chemical oscillation state is optically observable because the absorption spectrum of the solution depends on the ratio of reduced to oxidized catalyst (Fig. \ref{fig:SI_bz_stability}b).
In the reduced state the catalyst absorbs blue and green wavelengths \SIrange{450}{550}{\nano\metre}, so that one observes a red coloration of the solution. 
In the oxidized state the catalyst weakly absorbs red wavelengths around \SI{600}{\nano\metre}, leading to a faint blue coloration of the BZ solution.

\begin{figure*}
    \centering
    \includegraphics[width=0.5\textwidth]{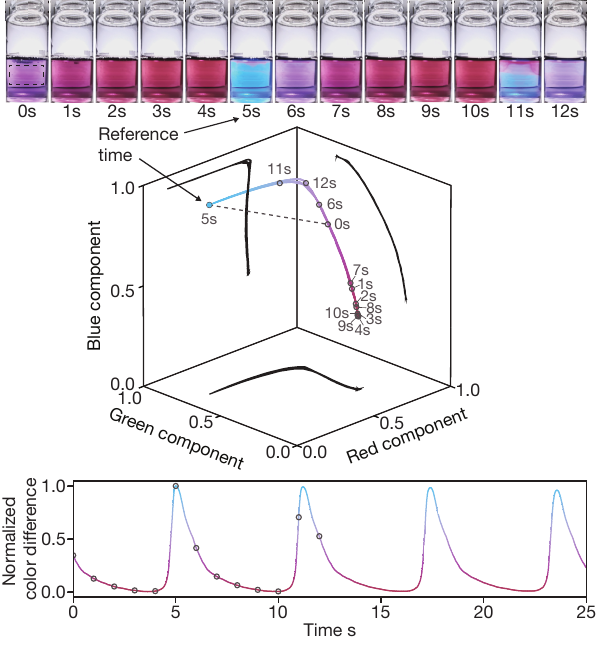}
    \caption{(Top) Snapshots of BZ reaction. (Middle) Average color in cropped region (dashed box top) plotted in RGB color space shows that the reaction follows a 1 dimensional curve in color space. (Bottom) A single trajectory is extracted from the BZ movies by first cropping the movies to a rectangle (dashed box top row) and then calulating the color difference between the average color of a frame and a reference frame (5s) using Euclidean distance in the Lab colorspace. The resulting trajectory is rescaled to lie between $[0, 1]$. }
    \label{fig:bz_trajectory_extraction}
\end{figure*}

A single trajectory was extracted from the movie~Fig.~\ref{fig:bz_trajectory_extraction}. First the movie was cropped to a rectangle containing only the BZ solution (dashed box). Next the frame with with the highest average blue channel value over the rectangle was chosen as a reference frame and the distance between this color and the average color was calculated using the Euclidean distance between the colors in the Lab colorspace. The distance was then normalized to lie in $[0, 1]$ and inverted so that the reference frame is a peak at $1$. \rv{Movies for four BZ experiments along with accompanying model fits learned by HDI are shown in the supplementary video file SI\_movie.mp4.}

A standard three-variable ODE model for the BZ reaction is the Oregonator model
\begin{subequations}\label{eq:oregonator}
\begin{align}
\dot{u} &= v - u\label{eq:oregu}\\
\dot{v} &= \frac{1}{\varepsilon_1}\Big(-w(v - \mu) - v^2 + v\Big) \label{eq:oregv}\\
\dot{w} &= \frac{1}{\varepsilon_2}\Big(fu + \phi - w(v + \mu)\Big)\label{eq:oregw}
\end{align}
\end{subequations}
where $u, v$, and $w$ correspond to the concentrations of the oxidized catalyst M$_\text{ox}$, bromous acid HBrO$_2$, and bromide Br$^-$ respectively. Here $\mu, f, \phi$ are all nonnegative. Usually we have that $\varepsilon_2 \ll \varepsilon_1$ so the inhibitor species $w$ can be \textit{adiabatically eliminated} by setting $fv + \phi - w(v + \mu) = 0$. Solving for $w$ we get that
\begin{equation}
    w = \frac{fu + \phi}{v + \mu}
\end{equation}
which leads to the two-component Oregonator model
\begin{subequations}\label{eq:oregonator_two}
\begin{align}
\dot{u} &= v - u\label{eq:oregtwou}\\
\dot{v} &= \frac{1}{\varepsilon_1}\Big(v(1 - v) - \frac{v - \mu}{v + \mu}(fu + \phi)\Big) \label{eq:oregtwov}
\end{align}
\end{subequations}

To build a two-variable polynomial HDI model that can match the dynamics of our experimental BZ recordings, we take inspiration from the two-component Oregonator model derived above. The first equation~\eqref{eq:oregtwou} in $\dot{u}$ is already of polynomial form and is in fact linear. The second equation ~\eqref{eq:oregtwov} in $\dot{v}$ has a non-polynomial rational term of the form $\frac{v - \mu}{v + \mu}$ which we must approximated through a polynomial expansion. A polynomial expansion is indeed possible because the coordinate trajectories of both Oregonator models~\eqref{eq:oregonator} and~\eqref{eq:oregonator_two} stay nonnegative with $v > \mu$ if initialized in this way. The Taylor series expansion of this rational function contains monomials of all integer degrees, and hence, we must decide where to truncate its polynomial expansion. In the region $v > \mu$, since $\mu \ll 1$ the rational function $\frac{v - \mu}{v + \mu}$ plateaus quickly to one, and hence, is well-approximated by a cubic polynomial in $v$. Therefore, the entire right hand side for the $\dot{v}$ equation~\eqref{eq:oregtwov} can be approximated by a quartic polynomial in $u, v$. To summarize, the two-component Oregonator model motivates us to search over all two-variable polynomial ODE models which are linear in their first equation and quartic in their second equation.

\begin{figure*}
    \centering
    \includegraphics[width=\textwidth]{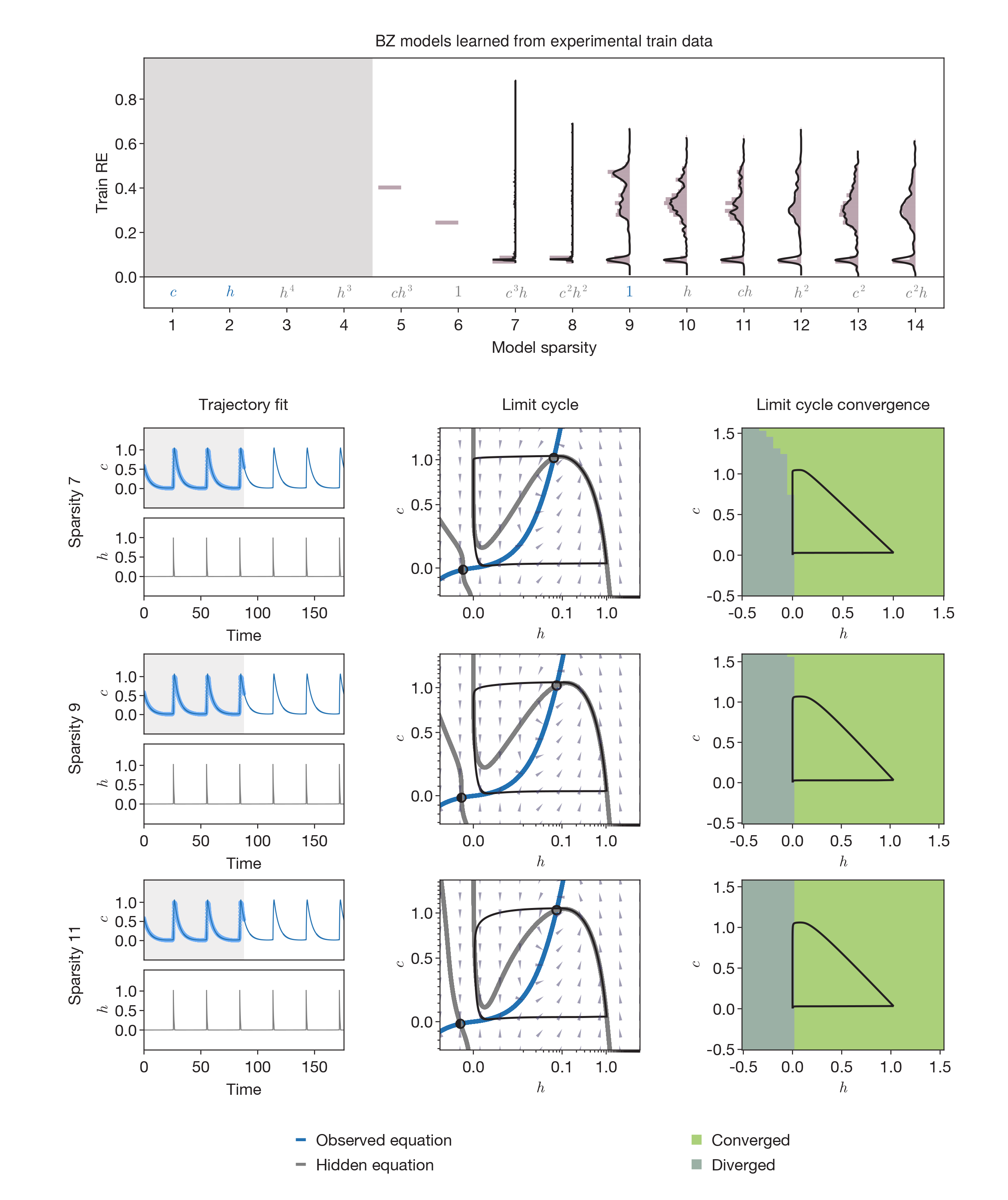}
    \caption{From three periods of the color change in a BZ chemical reaction, our method repeatedly fits two-variable ODE models with linear-quartic polynomials in $(v, h)$ to this \rv{training} data and learns a ranking of model terms shown in the top panel. At each sparsity level $s$, the top $s$ terms are combined into a new model and its coefficients are refit to the \rv{training} data from 2000 random parameter initializations. The losses on the \rv{training} data for all refit models of a given sparsity are plotted as a histogram in the top panel. The next three rows display the learned models with 7, 9, and 11 terms respectively. The first column displays the model fit to the data, the second column shows the \rv{stable} limit cycle, nullclines, fixed points, and vector fields of their phase plane dynamics, and the third columns marks in bright green the region of initial conditions in the phase plane that converge onto the limit cycle of each model.}
    \label{fig:bz_sparsity}
\end{figure*}

In Fig.~\ref{fig:bz_sparsity} we show the resulting sparse HDI models fit to 3 periods of the observed color $c$ of an experimental BZ chemical reaction. We plot how the histogram of model train losses change as more
terms are added into the model based on their ranked order found by the model term ranking step of our HDI sweep. All learned models with 7 or more terms display the correct nullcline and limit cycle dynamics as well as stability in the positive quadrant. 

\begin{figure*}
    \centering
    \includegraphics[width=\textwidth]{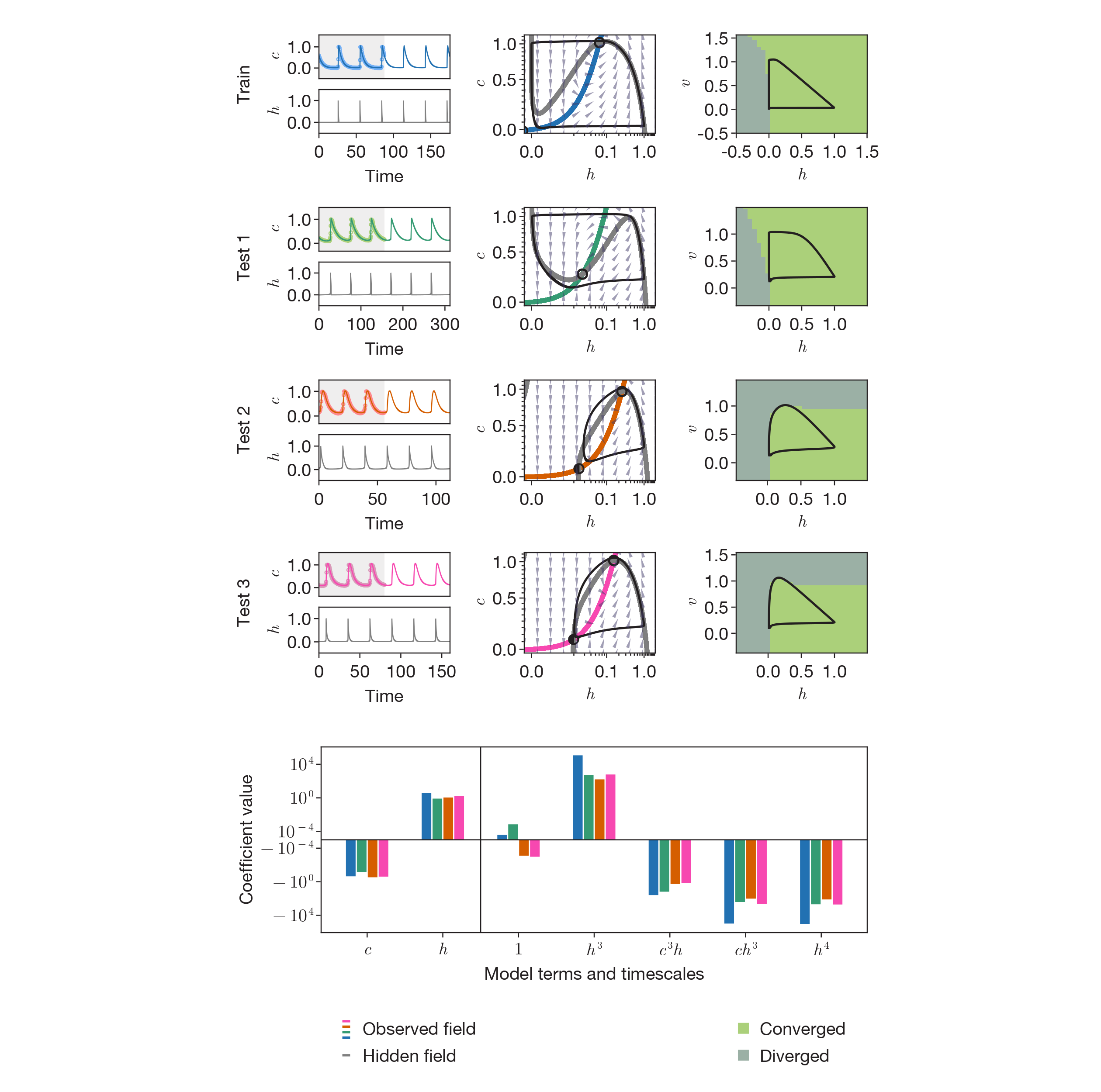}
    \caption{HDI model with seven terms fit to train recording of BZ experiment as well as three additional test recordings of new BZ experiments (first column). In column two, we show that models learned on \rv{training} data and first test recording have a stable limit cycle with the correct cubic-shaped nullcline for $\dot{h} = 0$ (gray line) along with the linear behavior of the $\dot{v} = 0$ nullcline (colored line). Learned models on second and third test examples also learn stable limit cycles albeit without the cubic nullcline behavior due to the increased width of the peaks in the experimental recordings. In the third column we show that models learned across all train and test datasets are sufficiently stable and converge to the learned limit cycle from most initial conditions (bright green squares in heatmap). Bottom plot shows that coefficients of the seven term model are consistent across all train and test recordings.}
    \label{fig:bz_train_and_test}
\end{figure*}

We show in Fig.~\ref{fig:bz_train_and_test} how the 7 term model generalizes to three additional recordings of BZ chemical reaction experiments.
\rv{The model is able to fit the new data, generated with different chemical concentrations, and moreover the coefficients remain consistent across experiments (Fig.~\ref{fig:bz_train_and_test}). It is interesting to consider how the coefficients depend on the chemical concentrations, and more generally how they would depend on some external system parameter, like concentration. Learning the parametric dependence of each of the 7 coefficients on the four varying chemical concentrations is not possible from four experiments, especially as this dependence is generically non-linear. 
Recent work~\cite{nicolaou2023datadriven} has demonstrated how, by introducing parametric terms into the model library within the SINDy framework, it is possible to learn parametric dependence of coefficients from numerical simulations of the Oregonator model. Following this approach, with more experimental recordings with systematically varied parameters, our framework could be extended to learn parametric dependencies of coefficients.}

\clearpage

\section{Chaotic Lorenz Attractor}
Here we study the dynamics of the chaotic Lorenz system when it is observed from its $x$-coordinate. The Lorenz system is given by the following equations
\begin{subequations}\label{eq:lorenz}
\begin{align}
\dot{x} &= \sigma (y - x) \label{eq:lorx}\\
\dot{y} &= x(\rho - z) - y \label{eq:lory}\\
\dot{z} &= xy - \beta z. \label{eq:lorz}
\end{align}
\end{subequations}
Below we show three experiments where we use our HDI pipeline to learn the Lorenz model from observations of only the $x$-coordinate, both $(x, y)$ coordinates, and $(x, y)$ coordinates with 10\% additive Gaussian noise.

\subsection{Observing Lorenz from a single $x$ coordinate}
First, let us consider the dynamics of $x$ alone. Taking~\eqref{eq:lorx} we can rewrite it as
\begin{align*}
    y = \frac{\dot{x}}{\sigma} + x, \quad \dot{y} = \frac{\ddot{x}}{\sigma} + \dot{x},
\end{align*}
then substituting this and~\eqref{eq:lorx} into~\eqref{eq:lory} gives us
\begin{align*}
    \frac{\ddot{x}}{\sigma} + \dot{x} = \rho x - xz - \frac{\dot{x}}{\sigma} - x \implies z = -\frac{1}{\sigma}\frac{\ddot{x}}{x} - \Big(1 + \frac{1}{\sigma}\Big)\frac{\dot{x}}{x} + \rho - 1.
\end{align*}
Given this expression for $z$ we can compute its derivative as
\begin{align*}
    \dot{z} = -\frac{1}{\sigma}\frac{\dddot{x}}{x} + \frac{1}{\sigma}\frac{\dot{x}\ddot{x}}{x^2} - \Big(1 + \frac{1}{\sigma}\Big)\frac{\ddot{x}}{x} + \Big(1 + \frac{1}{\sigma}\Big)\frac{\dot{x}^2}{x^2}.
\end{align*}
Finally, we can substitute the above two expressions for $y, z$, and $\dot{z}$ into~\eqref{eq:lorz} to write
\begin{align*}
    -\frac{1}{\sigma}\frac{\dddot{x}}{x} + \frac{1}{\sigma}\frac{\dot{x}\ddot{x}}{x^2} - \Big(1 + \frac{1}{\sigma}\Big)\frac{\ddot{x}}{x} + \Big(1 + \frac{1}{\sigma}\Big)\frac{\dot{x}^2}{x^2} = \frac{1}{\sigma}x\dot{x} + x^2 + \frac{\beta}{\sigma}\frac{\ddot{x}}{x} + \beta\Big(1 + \frac{1}{\sigma}\Big)\frac{\dot{x}}{x} - \beta(\rho - 1)
\end{align*}
which shows that
\begin{equation}\label{eq:lor_red_x}
    x\dddot{x} - \dot{x}\ddot{x} + (\sigma + \beta + 1)x\ddot{x} - (\sigma + 1)\dot{x}^2 + x^3\dot{x} + \beta(\sigma + 1)x\dot{x} + \sigma x^4 - \sigma\beta(\rho - 1)x^2 = 0.
\end{equation}

\rv{Reducing the Lorenz equations to a higher order ODE in solely the $x-$coordinate has been previously derived in~\cite{gouesbet1994global, linz1997nonlinear}.}

\begin{figure}[h]
    \centering
    \includegraphics{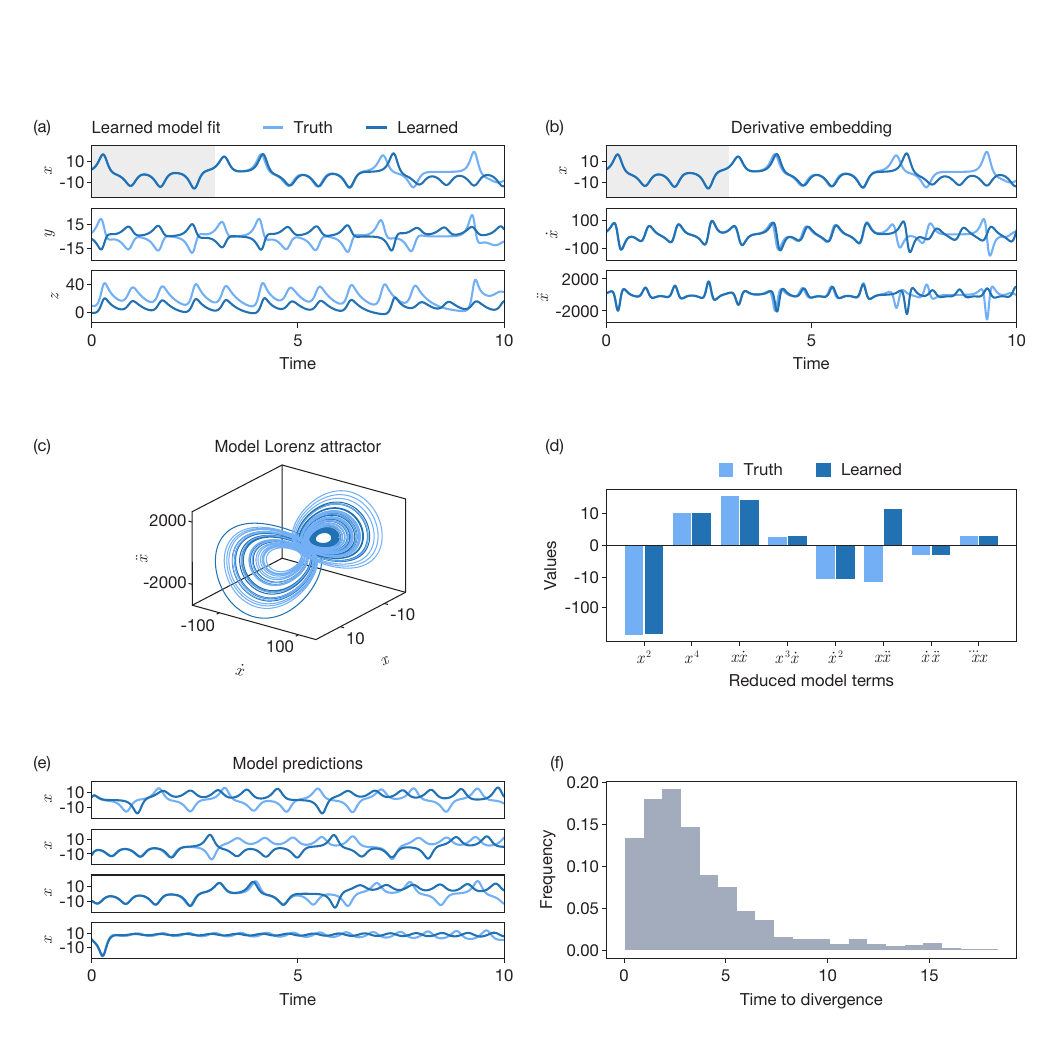}
    \caption{Lorenz model learned from HDI model sweep given only 3 time units (3 Lyapunov exponents) of observations of the $x$-coordinate from the true Lorenz system~\eqref{eq:lorenz} with parameters $\sigma = 10, \rho = 28, \beta = 8/3$. Learned and true Lorenz models in $(x, y, z)$ may not agree in unobserved $y, z$ coordinates and hence, are derivative embedded into $(x, \dot{x}, \ddot{x})$ coordinates where they can be fairly compared. (a) HDI model is fit to 3 time units of the $x$-coordinate simulated from a Lorenz attractor (\rv{training} data shown in gray). (b) Dynamics of learned model and true Lorenz system are derivative embedded into $(x, \dot{x}, \ddot{x})$ space. (c) Attractors of learned and true Lorenz systems in derivative space closely match. (d) Coefficients of reduced model for learned and true Lorenz system are in close agreement. (e) Learned model can predict $x$-coordinate of true Lorenz system when simulated from new initial conditions. (f) Histogram of time taken for learned model prediction to deviate from true Lorenz dynamics of $x$-coordinate for over 1000 initial conditions. On average, our learned model can predict the true Lorenz $x$ dynamics for 3.63 units of time (3.63 Lyapunov exponents) until it deviates from the true trajectory.}
    \label{fig:lorenz_1obs}
\end{figure}

Now we show how our model search procedure outlined in [Model Search Pipeline] determines a HDI model that accurately describes the Lorenz system. We give our pipeline 3 units of time (3 Lyapunov exponents) of just the $x$-coordinate of Lorenz simulated with parameters $\sigma = 10, \rho = 28, \beta = 8/3$ shown in the gray boxes of Fig.~\ref{fig:lorenz_1obs}(a). Following the model search procedure outlined above, we initialize 500 three-variable models in $(x, y, z)$ for 11 different sparsity parameters $\lambda$ for all polynomial ODE degree combinations up to cubic in all three equations. After one round of optimization, we keep those models which are chaotic (hence not divergent or periodic) and cluster them using hierarchical clustering at different cluster cutoff levels. Ranking the model terms by their coefficient of variation, our model sparsification step returns a HDI ODE with seven terms of the form
\begin{equation}\label{eq:lor_model_1obs}
\begin{aligned}
    \dot{x} &= p_1x + p_2y\\
    \dot{y} &= p_3x + p_4xz\\
    \dot{z} &= p_5 + p_6z + p_7xy
\end{aligned}
\end{equation}
which is strikingly similar to the true Lorenz system except a linear term in $y$ is missing from the $\dot{y}$ equation.

As shown in Fig.~\ref{fig:lorenz_1obs}(a) this ODE model exactly fits the \rv{training} data from the simulated Lorenz system (gray box) on which it was trained and is capable of predicting two additional branch switches of the Lorenz attractor which occur for 2 units of time (2 Lyapunov exponents) past the training window. The question becomes whether our model learned from just the $x$-coordinate of the Lorenz system exactly reduces to the true $x$-reduced Lorenz equation given in~\eqref{eq:lor_red_x}. However, using our automatic verification tool for model reductions described in section [Automatic tests for model reductions], we find that our model~\eqref{eq:lor_model_1obs} does not reduce to this form, no matter how we set the coefficients $p_1, \hdots, p_7$.

To see why this is, we reduce our learned model in the $x$-coordinate and obtain the following equation
\begin{equation}\label{eq:lor_model_1obs_red_x}
    x\dddot{x} - \dot{x}\ddot{x} + (p_1 + p_4p_6)x\ddot{x} + p_1\dot{x}^2 - p_4p_7x^3\dot{x} + p_1p_4p_6x\dot{x} + p_1p_4p_7x^4 + p_2(p_3p_6 - p_4p_5)x^2 = 0.
\end{equation}
The terms in this equation agree precisely with the $x$-reduced equation of the true Lorenz system in~\eqref{eq:lor_red_x}. Given that the true Lorenz system was simulated with parameters $\sigma = 10, \rho = 28, \beta = 8/3$, we now attempt to match the coefficient of our $x$-reduced model with that of Lorenz. This gives us the constraints
\begin{equation}\label{eq:red_constraints}
\begin{aligned}
    p_1 + p_4p_6 &= \sigma + \beta + 1\\
    p_1 &= -\sigma - 1\\
    p_4p_7 &= -1\\
    p_1p_4p_6 &= \beta(\sigma + 1)\\
    p_1p_4p_7 &= \sigma\\
    p_2(p_3p_6 - p_4p_5) &= -\sigma\beta(\rho - 1)
\end{aligned}
\end{equation}
from which we derive the contradiction that $p_1 = -\sigma$ and $p_1 = -\sigma - 1$. This explains why our automatic tool for model reductions could not find a set of coefficients $p_1, \hdots, p_7$ that would satisfy the true $x$-reduced Lorenz equation.

Despite this fact, our learned model closely matches the dynamics of the $x$-coordinate of the Lorenz system. Because the $(x, y, z)$ phase spaces of our learned model and the true Lorenz system only agree in the $x$ coordinate, we transform these models into the derivative embedding phase space $(x, \dot{x}, \ddot{x})$ using equations~\eqref{eq:lor_model_1obs} and~\eqref{eq:lorenz} respectively. The simulations of our learned and true models in this derivative embedded space are shown in Fig.~\ref{fig:lorenz_1obs}(b) and closely agree. In Fig.~\ref{fig:lorenz_1obs}(c) we further show that the attractors formed by these two models in derivative embedding space match almost exactly. This implies that the dynamics of our learned model in reduced model space $(x, \dot{x}, \ddot{x})$ given by~\eqref{eq:lor_model_1obs_red_x} closely matches the dynamics of the true reduced Lorenz equation in~\eqref{eq:lor_red_x}. By plotting the reduced model coefficients of our learned model against their true values in Fig.~\ref{fig:lorenz_1obs}(d), we confirm that this is indeed the case. Although the true reduced model coefficients can never be matched exactly as shown in~\eqref{eq:red_constraints}, our HDI model optimization has found the coefficient values $p_1, \hdots, p_7$ that are close to the true reduced Lorenz coefficients and reproduce the correct dynamics.

Finally, we show that our learned model can reliably predict the dynamics of the Lorenz $x$-coordinate by simulating our learned and true model from random initial conditions and studying the time at which the trajectories of both models diverge. To do this, we sample 1000 initial conditions in the derivative embedding space $(x, \dot{x}, \ddot{x}) \in [-20, 20] \times [-150, 150] \times [-2500, 2500]$ which are chosen uniformly at random from a box that contains the attractor shown in Fig.~\ref{fig:lorenz_1obs}(c). For each initial condition $(x, \dot{x}, \ddot{x})$ we invert equations~\eqref{eq:lorenz} and~\eqref{eq:lor_model_1obs} and find the points $(x, y, z)$ and $(x, y', z')$ in the phase spaces of both learned and true models respectively that correspond to this initial condition. The inversion of these dynamical systems is performed by solving a system of polynomial equations using homotopy continuation methods~\cite{HomotopyContinuation.jl}. Both the learned and true models are then simulated from their corresponding initial conditions $(x, y, z)$ and $(x, y', z')$ respectively. The $x$-coordinate trajectories of the learned and true Lorenz model are plotted for several random initial conditions $(x, \dot{x}, \ddot{x})$ in Fig.~\ref{fig:lorenz_1obs}(e). For each initial condition, we compute the first time $t^*$ when the learned and true models deviate in their $x$-coordinate by more than 4 units (e.g. 10\% of the interval $[-20, 20]$ in which $x$ lies). In Fig.~\ref{fig:lorenz_1obs}(f) we plot the time to divergence $t^*$ for 1000 uniformly sampled initial conditions $(x, \dot{x}, \ddot{x})$. We conclude that our learned model, trained on just the $x$-coordinate of Lorenz, can on average predict its dynamics for 3.63 units of time (3.63 Lyapunov exponents) from a random initial condition.

\subsection{Observing Lorenz in two $(x, y)$ coordinates}
Now let us study how our inference of the Lorenz system improve if we are allowed to observe both the $x$ and $y$ coordinates. First we show how the Lorenz system reduces when two coordinates $x$ and $y$ are observed whilst $z$ is hidden. Solving for $z$ in~\eqref{eq:lory} we find that $z = \rho - (\dot{y} + y)/x$ which implies that
\begin{equation}
    \dot{z} = -\frac{\ddot{y} + \dot{y}}{x} + \frac{\dot{x}\dot{y} + \dot{x}y}{x^2}.
\end{equation}
Substituting these expressions into~\eqref{eq:lorz} we obtain
\begin{equation}
    -\frac{\ddot{y} + \dot{y}}{x} + \frac{\dot{x}\dot{y} + \dot{x}y}{x^2} = xy - \rho\beta + \beta\frac{\dot{y} + y}{x}
\end{equation}
which multiplying by $x^2$ on both sides and rearranging gives us
\begin{equation}
    x\ddot{y} + x\dot{y} - \dot{x}\dot{y} - \dot{x}y + x^3y - \rho\beta x^2 + \beta x\dot{y} + \beta xy = 0.
\end{equation}
Hence, the reduced equations for Lorenz in $(x, y)$ read
\begin{equation}
\begin{gathered}
    \dot{x} = \sigma(y - x)\\
    x\ddot{y} - \dot{x}\dot{y} + (\beta + 1)x\dot{y} - \dot{x}y + x^3y - \rho\beta x^2 + \beta xy = 0.
\end{gathered}
\end{equation}
Note that the second expression is an implicit equation for $\ddot{y}$. To further simplify, we can remove the dependence on $\dot{x}$ in the second equation by substituting in $\dot{x} = \sigma(y - x)$ to finally get
\begin{equation}\label{eq:lor_red_xy}
\begin{gathered}
    \dot{x} = \sigma(y - x)\\
    x\ddot{y} + (\sigma + \beta + 1)x\dot{y} - \sigma y\dot{y} + x^3y - \rho\beta x^2 + (\sigma + \beta)xy - \sigma y^2 = 0.
\end{gathered}
\end{equation}
This makes it clear that $\dot{x}$ is a function of $(x, y)$ and that $\ddot{y}$ is an implicit function of $(x, y, \dot{y})$.

\begin{figure}[h!]
    \centering
    \includegraphics{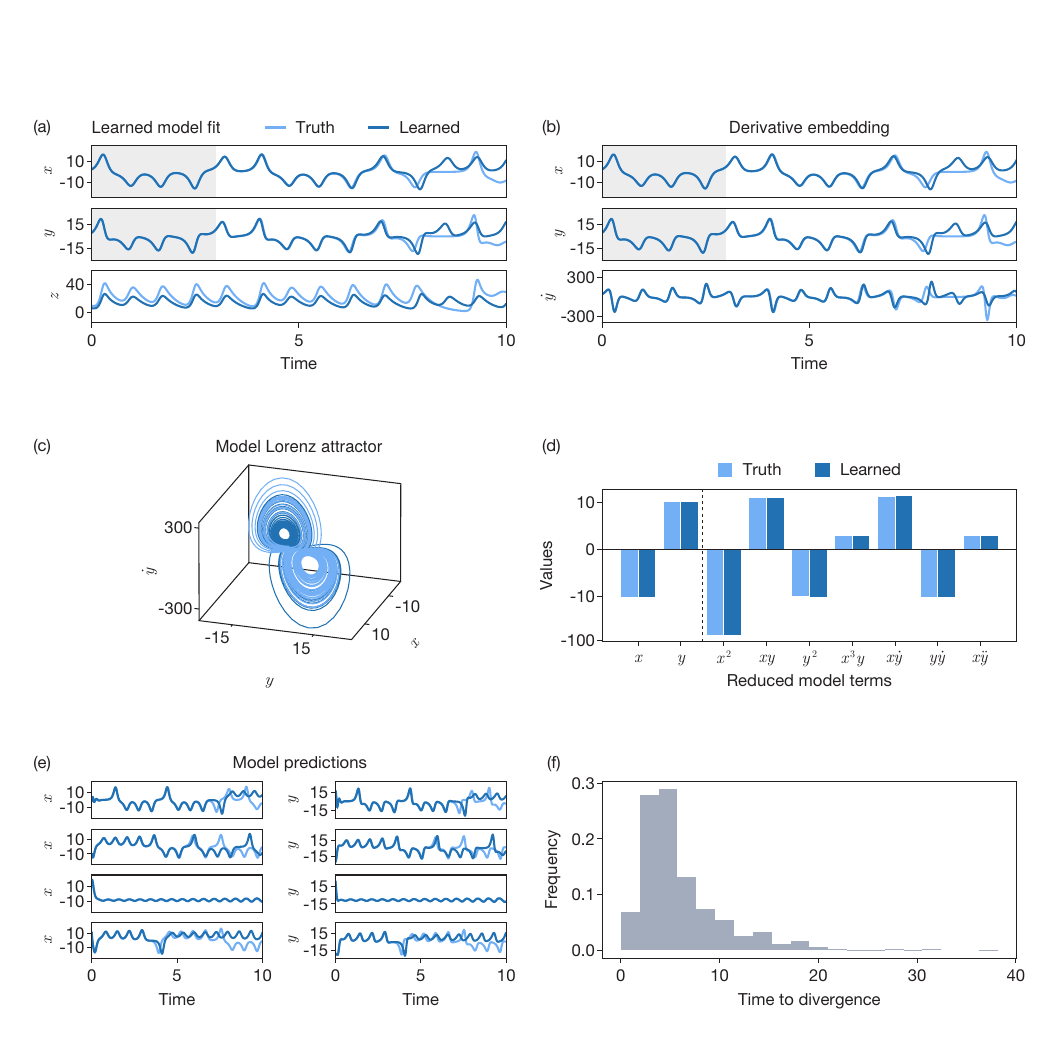}
    \caption{Lorenz model learned from HDI model sweep given only 3 time units of observations of the $x$ and $y$ coordinates from the true Lorenz system~\eqref{eq:lorenz} with parameters $\sigma = 10, \rho = 28, \beta = 8/3$. (a) HDI model trained on 3 time units of the $x$ and $y$ coordinates of Lorenz (gray boxes) can predict the dynamics for an additional 5 units of time and even learns correctly the unobserved $z$-coordinate dynamics. (b) Learned and true Lorenz models also agree when embedded in the derivative phase space $(x, y, \dot{y})$. (c) Limit cycles of learned and true models match exactly in derivative embedding space. (d) Coefficients of learned~\eqref{eq:lor_model_2obs_red_xy} and true~\eqref{eq:lor_red_xy} reduced models agree exactly. (e) Dynamics of learned and true Lorenz models in $(x, y)$ coordinates have the same branch-switching dynamics and agree for the first 3-10 units of time when simulated form random initial conditions. (f) Histogram of time length for which learned model trajectory agrees (predicts) the true dynamics of Lorenz from a random initial condition before the two trajectories diverge. Our learned model can predict the dynamics of the true Lorenz system for 6 units of time on average.}
    \label{fig:lorenz_2obs}
\end{figure}

Now we show how our model search procedure outlined in [Model Search Pipeline] is able to discover the equations of the Lorenz system from observations of the $x$ and $y$ coordinates. We begin as before by simulating the Lorenz system~\eqref{eq:lorenz} with parameters $\sigma = 10, \rho = 28, \beta = 8/3$ for 3 units of time and only save the trajectories of the $x$ and $y$ coordinates highlighted in gray boxes in Fig.~\ref{fig:lorenz_2obs}(a). On this \rv{training} data of just two observable trajectories, we optimize 500 three-variable HDI models in $(x, y, z)$ for 11 different sparsity parameters $\lambda$ for all polynomial ODE degree combinations up to cubic in all three equations. We then remove all models which are not chaotic (hence not divergent or periodic) and cluster the remaining models using hierarchical clustering. The model terms are then sorted from most to least important by an aggregated ranking algorithm described in the Model Term Ranking step of our pipeline. Finally the Model Sparsification step keeps the top seven terms in our ranking which correspond exactly to the true Lorenz equations
\begin{equation}\label{eq:lor_model_2obs}
\begin{aligned}
    \dot{x} &= p_1x + p_2y\\
    \dot{y} &= p_3x + p_4y + p_5xz\\
    \dot{z} &= p_6z + p_7xy \, .
\end{aligned}
\end{equation}
Given that this model has exactly the same terms as the Lorenz system, it reduces to the correct dynamics in $(x, y)$ of the form
\begin{equation}\label{eq:lor_model_2obs_red_xy}
\begin{gathered}
    \dot{x} = p_1x + p_2y\\
    x\ddot{y} - (p_1 + p_4 + p_6)x\dot{y} - p_2y\dot{y} - p_5p_7x^3y + p_3p_6x^2 + p_4(p_1 + p_6)xy + p_2p_4y^2 = 0
\end{gathered}
\end{equation}
In Fig.~\ref{fig:lorenz_2obs}(a) we show the fit of our learned model~\eqref{eq:lor_model_2obs} to the $(x, y)$ coordinates of Lorenz (\rv{training} data in gray boxes) and find that our model can predict the dynamics of Lorenz for an additional 5 units of time past the training data. Because the $z$ coordinate of Lorenz is unobserved, our learned model does not match it exactly, but in this case, it can be scaled properly so that they match. Alternatively, in order to compare our learned and true Lorenz models, we can transform them into the $(x, y, \dot{y})$ phase space of their reduced models~\eqref{eq:lor_red_xy} and~\eqref{eq:lor_model_2obs_red_xy}. In this derivative embedded phase space, we plot the dynamics of both models in Fig.~\ref{fig:lorenz_2obs}(b) and also show close agreement of their limit cycle dynamics in Fig.~\ref{fig:lorenz_2obs}(c). To further convince ourselves that our HDI model search has learned the correct model, we plot the $(x, y)$-reduced coefficients of our learned model from~\eqref{eq:lor_model_2obs_red_xy} against the true reduced coefficients of Lorenz in~\eqref{eq:lor_red_xy} where we see exact agreement.

Lastly, we study the predictive ability of our learned model by simulating it and the true Lorenz system from 1000 initial conditions $(x, y, \dot{y}) \in [-20, 20] \times [-30, 30] \times [-400, 400]$ chosen uniformly at random from a box in the derivative embedded phase space where both models can be compared. In Fig.~\ref{fig:lorenz_2obs}(e) we plot simulations of the $(x, y)$ coordinates of the learned and true models from several randomly chosen initial conditions $(x, y, \dot{y})$. For each simulation from a random initial condition, we compute the first time $t^*$ when the difference $(\Delta x, \Delta y)$ of the $(x, y)$-coordinates of the learned and true models exceed the condition
\begin{equation}
    \sqrt{\Big(\frac{\Delta x}{40}\Big)^2 + \Big(\frac{\Delta y}{60}\Big)^2} < 0.1.
\end{equation}
This condition tests whether the difference between the $(x, y)$ dynamics of the learned and true model normalized to the unit square exceed 10\% of the unit square sidelength (e.g. exceed 0.1). Over 1000 random initial conditions $(x, y, \dot{y})$ chosen from the box, we plot the divergence time $t^*$ in a histogram in Fig.~\ref{fig:lorenz_2obs}(f). Finally, we find that our correctly learned Lorenz model is predictive for an average of 6 time units (6 Lyapunov exponents). In comparison with our Lorenz model learned from just the $x$-coordinate, our new Lorenz model learned from both $x, y$ coordinates can predict the dynamics of Lorenz for twice as long on average.

\subsection{\rv{Observing Lorenz in two $(x, y)$ coordinates with noise}}

\rv{Here we repeat the same analysis as above but add 10\% Gaussian noise to the two $(x, y)$ coordinates that are observed. Performing our HDI pipeline, the Model Term Ranking and Model Sparsification steps keep the top 8 terms
\begin{equation}\label{eq:lor_model_2obs_noise}
\begin{aligned}
    \dot{x} &= p_1x + p_2y\\
    \dot{y} &= p_3x + p_4y + p_5xz\\
    \dot{z} &= p_6 + p_7z + p_8xy \, .
\end{aligned}
\end{equation}
The model above is almost identical to the true Lorenz system except for the addition of a bias term $p_6$ in the equation for $\dot{z}$. This addition of a bias term is a result of our model learning a translated version of the coordinate $z \mapsto z + c$. Specifically, this translation came from the fact that we centered all models in their $z$-coordinate in order to improve the Model Clustering step of our HDI pipeline whereas the true Lorenz model $z$-coordinate has nonzero mean. This centering step was necessary as each learned model in our sample had too much variability in its translation of the $z$-coordinate which negatively affected the model clustering.}

\begin{figure}[h!]
    \centering
    \includegraphics{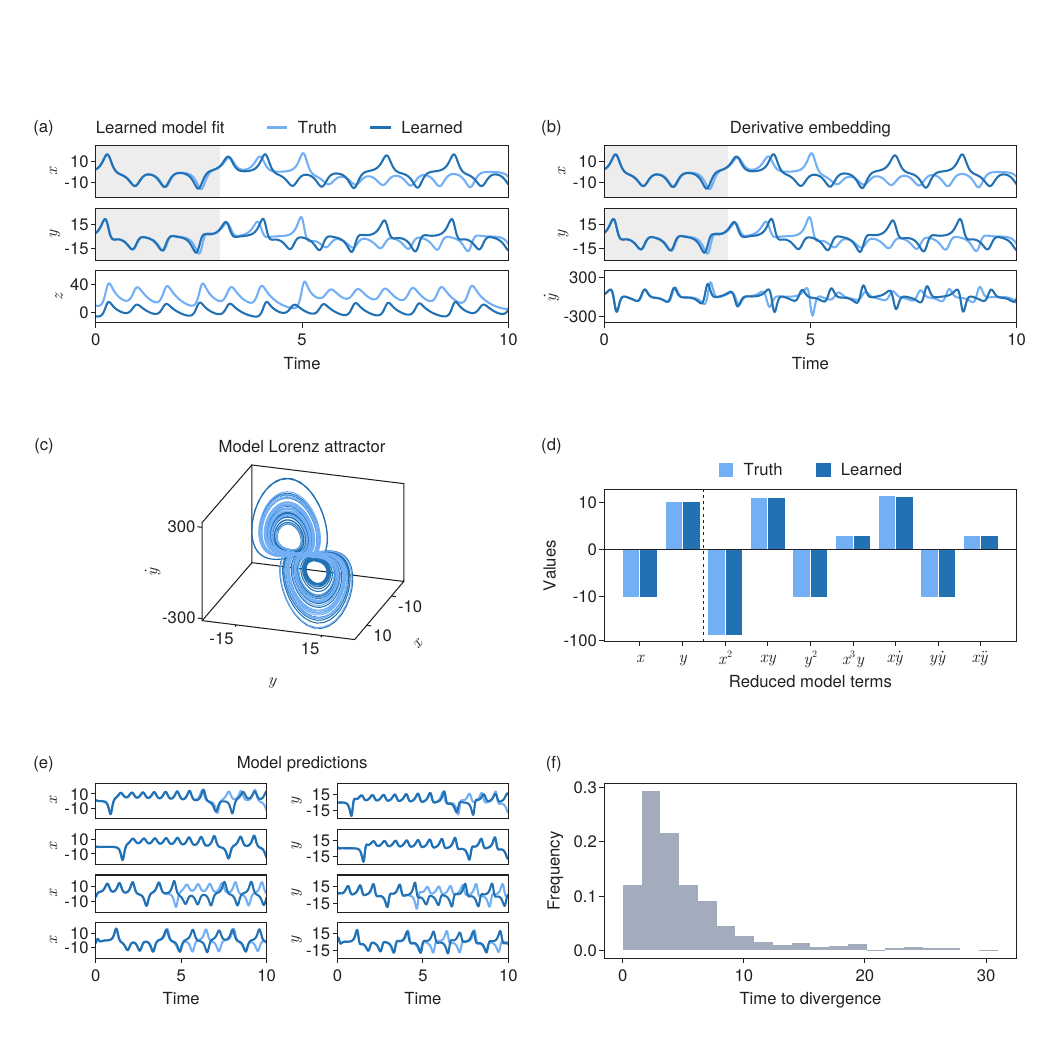}
    \caption{\rv{Noisy Lorenz model learned from HDI model sweep given only 3 time units of observations of the $x$ and $y$ coordinates with 10\% Gaussian additive noise from the true Lorenz system~\eqref{eq:lorenz} with parameters $\sigma = 10, \rho = 28, \beta = 8/3$. (a) HDI model trained on 3 time units of the $x$ and $y$ coordinates of Lorenz (gray boxes) can predict the dynamics for an additional 5 units of time and learns a similar dynamics in its $z$-coordinate which was unobserved. The $z$-coordinate of our learned model is centered to have mean zero. (b) Learned and true Lorenz models also agree when embedded in the derivative phase space $(x, y, \dot{y})$. (c) Limit cycles of learned and true models match exactly in derivative embedding space. (d) Under 10\% additive Gaussian noise for observed data, the Coefficients of learned~\eqref{eq:lor_model_2obs_red_xy_noise} and true~\eqref{eq:lor_red_xy} reduced models agree exactly. (e) Dynamics of learned and true Lorenz models in $(x, y)$ coordinates have the same branch-switching dynamics and agree for approximately the first 5 units of time when simulated form random initial conditions. (f) Histogram of time length for which learned model trajectory agrees (predicts) the true dynamics of Lorenz from a random initial condition before the two trajectories diverge. Our learned model can predict the dynamics of the true Lorenz system for 5 units of time on average.}}
    \label{fig:lorenz_2obs_noise}
\end{figure}

\rv{For the learned model structure in~\eqref{eq:lor_model_2obs_noise}, its reduced equation in the coordinates $(x, y, \dot{y})$ is given by
\begin{equation}\label{eq:lor_model_2obs_red_xy_noise}
\begin{gathered}
    \dot{x} = p_1x + p_2y\\
    x\ddot{y} - (p_1 + p_4 + p_7)x\dot{y} - p_2y\dot{y} - p_5p_8x^3y + (p_3p_7 - p_5p_6)x^2 + p_4(p_1 + p_7)xy + p_2p_4y^2 = 0
\end{gathered}
\end{equation}
which has the same exact terms as the true reduced Lorenz equation from~\eqref{eq:lor_model_2obs_red_xy}. In Fig.~\ref{fig:lorenz_2obs_noise} we see that on 10\% noised observations of the $(x, y)$ coordinates, the model learned by HDI has exactly the correct reduced model coefficients, obtains the correct chaotic attractor shape, and is predictive for 1-2 branch switches of the Lorenz attractor.}

\newpage

\rv{\section{Extension to non-polynomial basis functions}}

\rv{The methodology outlined above is readily extended to other basis functions such as trigonometric polynomials. The same MSE objective function~\ref{eq:obj} and optimization procedure is used but with a modified ODE system,
\begin{equation} \label{eq:trig_oscillator}
\dot{\theta}_i = \omega_i + \sum_{j = i - 1}^{i + 1} A_{ij}\cos(b_{ij} \theta_j + c_{ij})
\end{equation}
where $1 
\le i \le M$ and the index $i$ is defined cyclically so that $i = M + 1 = 1$ and $i = -1 = M$. The first $m$ fields are assumed to be observed.}

\begin{figure}
\centering\includegraphics[width = \textwidth]{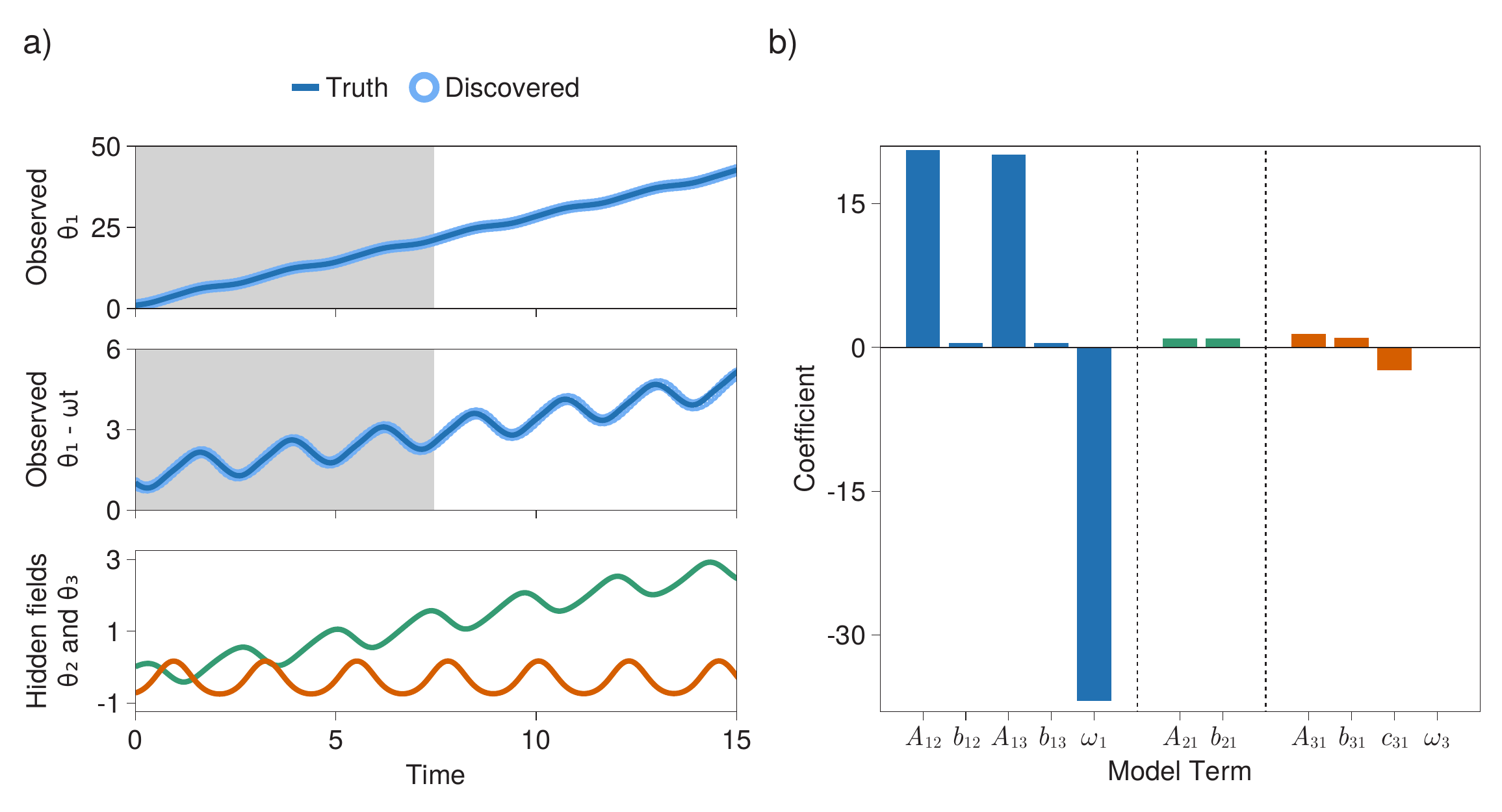}
    \caption{\rv{HDI discovers predictive model using a trigonometric basis. a) Discovered model (solid line) accurately reproduces the observed data (open circles, grey box) and is predictive over a window of the same length as observed data (top). Data with linear trend $\omega t$ subtracted makes flucuations clearer (middle). Hidded fields discovered by the model (bottom). b) Sparse set of coefficients discovered by HDI.}\label{fig:adler_results}}
\end{figure}

\rv{To test HDI on a non-polynomial basis we simulate~\eqref{eq:trig_oscillator} with $\omega_i = 2.5$, $A_{ii} = 0.5$ and $A_{ij} = (-1)^i$ for $i \ne j$ and $M = 3$. Using HDI on a single observed variable $m = 1$ we learn a sparse model of the form~\eqref{eq:trig_oscillator}. The optimization procedure is repeated $200$ times with $\lambda = 10^{-5}$. The model with the lowest data error is chosen for sparsification. Terms with a magnitude larger that $10^{-10}$ are retained. The coefficients of the resulting sparsity pattern are then reoptimized with no sparsity regularization. The HDI framework is able to discover a model that matches the observed data accurately and is predictive over the same length of time as training data window~(Fig.~\ref{fig:adler_results})}~\\

\section{Automatic Tests for Model Reductions}
In the examples above, we meticulously derived the reduced forms for several two and three-variable dynamical systems which could only be observed in a subset of their variables. This allowed us to verify for example that the models we learned on FHN through our HDI pipeline were truly correct in the sense that they could reproduce all possible dynamics of the observed $v$ coordinate. Generally, our model search returns an $M$-variable dynamical system in $(x_1, \hdots, x_m, h_{m+1}, \hdots, h_M)$ with parameter vector $\mathbf{w} \in \mathbb{R}^l$ of the form
\begin{equation}\label{eq:hfm2}
\begin{aligned}
    \dot{x}_k &= f_k(x_1, \hdots, h_M; \mathbf{w}), \quad 1 \leq k \leq m\\
    \dot{h}_k &= f_k(x_1, \hdots, h_M; \mathbf{w}), \quad m < k \leq M
\end{aligned}
\end{equation}
where we can only observe the first $m$ variables $x_1, \hdots, x_m$. Here $\mathbf{w}$ is the vector of parameters/coefficients which are optimized in order for the model to fit the observed data. For example, all HDI models we optimize in this paper take the polynomial form
\begin{equation}
\begin{aligned}
    \dot{x}_k &= \sum_{\bm\alpha \in S_k} w_{\bm\alpha}^k x_1^{\alpha_1}\hdots h_M^{\alpha_M}, \quad 1 \leq k \leq m\\
    \dot{h}_k &= \sum_{\bm\alpha \in S_k} w_{\bm\alpha}^k x_1^{\alpha_1} \hdots h_M^{\alpha_M}, \quad m < k \leq M
\end{aligned}
\end{equation}
and are parameterized by the weight vector $\mathbf{w} = \{w_{\bm\alpha}^k: \bm\alpha \in S_k, \ 1 \leq k \leq M\} \in \mathbb{R}^l$ with length $l = |S_1| + \hdots + |S_M|$ where each set $S_k \subset \mathbb{N}^M$ for equation $k$ indicates the polynomial terms (monomial degrees) present in the right hand side of that equation.

We would like to automatically check whether our learned model~\eqref{eq:hfm2} satisfies a known reduced model
\begin{equation}\label{eq:genred}
    g_k(x_k^{d_k}, \{x_1^{(d)}\}_{d=0}^{d_1-1}, \{x_2^{(d)}\}_{d=0}^{d_2-1}, \hdots, \{x_m^{(d)}\}_{d=0}^{d_m-1}) = 0, \quad 1 \leq k \leq m.
\end{equation}
The reduced model above is written in its most general form where the $k$th equation of the reduced model implicitly relates the $d_k$th derivative of the observed variable $x_k$ to all lower order derivatives of the observed variables $x_1, \hdots, x_m$ using the functional constraint $g_k \equiv 0$. This reduced model has degree $(d_1, \hdots, d_m)$ because $d_k$ is the largest derivative of each observed variable $x_k$ for $1 \leq k \leq m$.

For example, if we observe the FHN system
\begin{equation}
\begin{aligned}
    \dot{v} &= v - cv^3 - w + I\\
    \dot{w} &= \frac{1}{\tau}\Big(v + a - bw\Big)
\end{aligned}
\end{equation}
in only its $v$ coordinate, then its dynamics in $v$ is given by a reduced model of degree 2 of the form
\begin{equation}
    g(\ddot{v}, \dot{v}, v) = 0
\end{equation}
where
\begin{equation}\label{eq:genred_fhn}
    g(\ddot{v}, \dot{v}, v) = \ddot{v} + 3cv^2\dot{v} - \Big(1 - \frac{b}{\tau}\Big)\dot{v} + \frac{bc}{\tau}v^3 - \frac{b-1}{\tau}v - \frac{bI - a}{\tau}.
\end{equation} 

As another example, if we observe the Lorenz system
\begin{equation}
\begin{aligned}
\dot{x} &= \sigma (y - x)\\
\dot{y} &= x(\rho - z) - y\\
\dot{z} &= xy - \beta z
\end{aligned}
\end{equation}
only in the $x$ and $y$ coordinates, then its reduced model has the form
\begin{equation}
\begin{gathered}
    g_1(\dot{x}, x, y) = 0\\
    g_2(\ddot{y}, x, y, \dot{y}) = 0
\end{gathered}
\end{equation}
where
\begin{equation}\label{eq:genred_lorenz}
\begin{gathered}
    g_1(\dot{x}, x, y) = \dot{x} + \sigma x - \sigma y\\
    g_2(\ddot{y}, x, y, \dot{y}) = x\ddot{y} + (\sigma + \beta + 1)x\dot{y} - \sigma y\dot{y} + x^3y - \rho\beta x^2 + (\sigma + \beta)xy - \sigma y^2.
\end{gathered}
\end{equation}

If we are given the dynamical system given in~\eqref{eq:hfm2}, we would like to understand whether its observed coordinates satisfy the reduced equation in~\eqref{eq:genred}. To do so, we first write out the higher order derivatives of each observed coordinate $x_k, \dot{x}_k, \ddot{x}_k, \hdots, x_k^{(d_k)}$ as functions of the $M$ model variables $(x_1, \hdots, x_m, h_{m+1}, \hdots, h_M)$ and the model parameters $\mathbf{w}$. This can be simply done by repeatedly differentiating the $k$th equation in~\eqref{eq:hfm2}. Now that we have expressed the higher order derivative of every observed coordinate
\begin{equation}
    x_k^{(d)} := x_k^{(d)}(x_1, \hdots, x_m, h_{m+1}, \hdots, h_M; \mathbf{w}), \quad 1 \leq d \leq d_k, \quad 1 \leq k \leq m
\end{equation}
as a function of the original model variables and parameters, we can then substitute them back into each equation in~\eqref{eq:genred} to obtain the set of constraints
\begin{equation}\label{eq:genred_sub}
    \overline{g}_k(x_1, \hdots, h_M; \mathbf{w}) := g_k(x_k^{d_k}(x_1, \hdots, h_M; \mathbf{w}), \{x_1^{(d)}(x_1, \hdots, h_M; \mathbf{w})\}_{d=0}^{d_1-1}, \hdots, \{x_m^{(d)}(x_1, \hdots, h_M; \mathbf{w})\}_{d=0}^{d_m-1}) = 0
\end{equation}
for all $1 \leq k \leq m$ and $(x_1, \hdots, h_M) \in \mathbb{R}^M$.

If the original dynamical system~\eqref{eq:hfm2} has right hand side functions $f_k$ which are polynomials of $x_1, \hdots, h_M$ and the entries of $\mathbf{w}$, then it usually holds that the implicit constraints $g_1, \hdots g_m$ of the reduced model are also polynomials in these variables as shown in the examples of FHN and Lorenz above. In fact, this has been shown to hold for all of the polynomial dynamical systems considered in this paper. When the right hand side functions $f_1, \hdots, f_m$ are polynomial functions of $x_1, \hdots, h_M, \mathbf{w}$, then it is easy to see that the higher order derivatives $x_k^{(d)}$ of the observed variables are also polynomial functions of $x_1, \hdots, h_M, \mathbf{w}$. Because the implicit constraints $g_1, \hdots g_m$ of the reduced equation are also of polynomial form, this implies that the functions $\overline{g}_k$ are polynomials of $x_1, \hdots, h_M$ and the entries of $\mathbf{w}$.

Let us demonstrate this on the example of the R\"{o}ssler system,
\begin{equation}\label{eq:rossler}
\begin{aligned}
\dot{x} &= -y - z\\
\dot{y} &= rx + ay\\
\dot{z} &= b + z(x - c).
\end{aligned}
\end{equation}
If we observe this system in the $x$ and $y$ coordinates (with $z$ hidden) it is not hard to check that its reduced model takes the form
\begin{equation}
\begin{gathered}
    \ddot{x} = x\dot{x} - c\dot{x} + xy - rx - (a + c)y - b\\
    \dot{y} = rx + ay.
\end{gathered}
\end{equation}
In our notation, this reduced model can be rewritten as
\begin{equation}\label{eq:genred_rossler}
\begin{gathered}
    g_1(\ddot{x}, x, \dot{x}, y) := \ddot{x} - x\dot{x} + c\dot{x} - xy + rx + (a + c)y + b = 0\\
    g_2(\dot{y}, x, y) = \dot{y} - rx - ay = 0.
\end{gathered}
\end{equation}
Note that because our original R\"{o}ssler equations were of polynomial form, the implicit constraints $g_1, g_2$ of the reduced model are also polynomial functions. Now suppose that the parametrized model we have learned through our model search has the same sparsity pattern as the R\"{o}ssler equations
\begin{equation}\label{eq:learned_rossler}
\begin{aligned}
\dot{x} &= f_1(x, y, z; \mathbf{w}) := w_1y + w_2z\\
\dot{y} &= f_2(x, y, z; \mathbf{w}) := w_3x + w_4y\\
\dot{z} &= f_3(x, y, z; \mathbf{w}) := w_5 + w_6z + w_7xz
\end{aligned}
\end{equation}
where $\mathbf{w} = (w_1, \hdots, w_7)^T$. In general, the learned model could take any form, but here for simplicity we choose it to agree with the form of the R\"{o}ssler equations. Because our reduced model in~\eqref{eq:genred_rossler} depends on $x, y, \dot{x}, \dot{y}, \ddot{x}$ we use the equations in~\eqref{eq:learned_rossler} to write
\begin{equation}\label{eq:learned_rossler_higher_order}
\begin{aligned}
    \dot{x}(x, y, z; \mathbf{w}) &= w_1y + w_2z\\
    \dot{y}(x, y, z; \mathbf{w}) &= w_3x + w_4y\\
    \ddot{x}(x, y, z; \mathbf{w}) &= w_1\dot{y} + w_2\dot{z} = w_2w_7xz + w_2w_6z + w_1w_4y + w_1w_3x + w_2w_5
\end{aligned}
\end{equation}
As mentioned above, $\dot{x}, \dot{y}, \ddot{x}$ are all polynomial functions of $x, y, z$ and the entries of $\mathbf{w}$ because our original learned model~\eqref{eq:learned_rossler} was of polynomial form.

Now that we have expressed the higher order derivatives of $x$ and $y$ in the original phase space coordinates $(x, y, z)$ and parameters $\mathbf{w}$ of our learned model, we can substitute them into our true reduced R\"{o}ssler model~\eqref{eq:genred_rossler} to get
\begin{equation}
\begin{aligned}
    \overline{g}_1(x, y, z; \mathbf{w}) &:= -(w_1 + 1)xy + (w_2w_7 - w_2)xz + (w_1w_3 + r)x\\
    &\qquad+ (w_1w_4 + cw_1 + a + c)y + (w_2w_6 + cw_2)z + (w_2w_5 + b) = 0\\
    \overline{g}_2(x, y, z; \mathbf{w}) &:= (w_3 - r)x + (w_4 - a)y = 0.
\end{aligned}
\end{equation}
Because we have composed the polynomial functions in equations~\eqref{eq:learned_rossler} and~\eqref{eq:learned_rossler_higher_order}, this implies that the reduced model constraints $\overline{g}_1, \overline{g}_2$ are also polynomial functions of $x, y, z$ and the entries of $\mathbf{w}$. This completes our example and shows how to rewrite the reduced model constraints $g_k$ of the R\"{o}ssler system~\eqref{eq:rossler} as functions of the variables and parameters of a learned model~\eqref{eq:learned_rossler}. This exact same procedure can be applied to study models learned on the FHN and Lorenz systems above.

Finally, in order to show that a general dynamical system~\eqref{eq:hfm2} with learned parameters $\mathbf{w}$ exactly reduces to~\eqref{eq:genred}, we need to prove that each function $\overline{g}_k: \mathbb{R}^M \to \mathbb{R}$ with fixed parameters $\mathbf{w}$ is equal to zero for all possible inputs $x_1, \hdots, h_M$. Of course, due to noise in the data or fluctuations in the model optimization, the model parameters $\mathbf{w}$ learned during the course of optimization cannot perfectly set $\overline{g}_k$ to be the zero function. For example, our learned models of FHN and Lorenz correctly discovered the exact polynomial forms of the true reduced models in~\eqref{eq:genred_fhn} and~\eqref{eq:genred_lorenz} but the coefficients of these learned reduced models fluctuate around the true values. Hence, our learned HDI models on the FHN and Lorenz systems would not be able to pass the unreasonably strict requirements that the true reduced model is satisfied perfectly (i.e. $\overline{g}_k(\cdot; \mathbf{w}) \equiv 0$).

A much more useful way to test if a dynamical model reduces to the correct form is to ask whether there is a way to \textit{perturb} its parameters $\mathbf{w}$ such that it exactly satisfies the true reduced model. Namely, in order to prove that our learned model~\eqref{eq:hfm2} can reduce to the correct form in~\eqref{eq:genred}, we need to show that there exists a set of parameters $\mathbf{w}^* \in \mathbb{R}^l$ such that the reduced model constraints in~\eqref{eq:genred_sub} are satisfied exactly
\begin{equation}\label{eq:genred_sub_opt}
    \overline{g}_k(x_1, \hdots, h_M; \mathbf{w}^*) = 0, \quad \forall \ 1 \leq k \leq m, \ (x_1, \hdots, h_M) \in \mathbb{R}^M.
\end{equation}

If there is no prior knowledge about the functional form of the reduced model constraints $\overline{g}_k$, then we could find $\mathbf{w}^*$ by sampling each function $\overline{g}_1, \hdots, \overline{g}_m$ at different points in phase space $(x_1, \hdots, h_M) \in \mathbb{R}^M$ and using Newton's method or general nonlinear optimization to find the optimal parameter vector $\mathbf{w}^*$ that sets all $m$ functions to zero at the sample points. If such a $\mathbf{w}^*$ does not exist, this would prove that our learned dynamical system does not match the correct reduced model.

As seen above on the example of the R\"{o}ssler system, when learning dynamical systems of  polynomial form, the reduced model constraints $\overline{g}_1, \hdots, \overline{g}_m$ become polynomial functions of the variables $x_1, \hdots, h_M$ and model parameters $\mathbf{w} \in \mathbb{R}^l$. Namely, we can write them out as
\begin{equation}
    \overline{g}_k(x_1, \hdots, h_M; \mathbf{w}) = \sum_{\bm{\alpha} \in I_k} p_k^{\bm{\alpha}}(\mathbf{w})x_1^{\alpha_1}\hdots h_M^{\alpha_M}, \quad 1 \leq k \leq m
\end{equation}
where $p_k^{\bm{\alpha}}: \mathbb{R}^l \to \mathbb{R}$ is a polynomial and $I_k \subset \mathbb{N}^M$ is the set of monomial degrees appearing in the right hand side of $\overline{g}_k$. In expressing the polynomial $\overline{g}_k$ above, we have simply isolated each monomial term in $x_1, \hdots, h_M$ as a separate term in the sum.
Now, we can find the optimal value of $\mathbf{w}^*$ that analytically sets these constraints to zero $\overline{g}_k(\cdot; \mathbf{w}) \equiv 0$ by solving a polynomial system of equations
\begin{equation}
    p_k^{\bm{\alpha}}(\mathbf{w}) = 0, \quad \forall \ \bm{\alpha} \in I_k, \ 1 \leq k \leq m.
\end{equation}
This system of polynomial equations, which is usually overdetermined, can be solved either by hand or with homotopy continuation methods~\cite{HomotopyContinuation.jl} to find the model parameters $\mathbf{w}^*$ that analytically set these constraints to zero and hence, satisfy the true reduced model. It is important to note that the solution $\mathbf{w}^*$ may not exist, in which case the learned dynamical system does not reduce to the correct model. In other cases, the optimal solution $\mathbf{w}^*$ may be unique, may have a discrete number of solutions, or may even have a continuous set of solutions. It is up to the user to specify which solutions of $\mathbf{w}^*$ they are willing to consider (real, complex, nonsingular, etc.) We summarize this procedure for testing model reductions in the following algorithm:
\IncMargin{1em}
\begin{algorithm}[!t]
\SetKwData{Left}{left}\SetKwData{This}{this}\SetKwData{Up}{up}
\SetKwFunction{Union}{Union}\SetKwFunction{FindCompress}{FindCompress}
\SetKwInOut{Input}{input}\SetKwInOut{Output}{output}
\Input{A learned polynomial HDI model
\begin{equation}\label{eq:hfm3}
\begin{aligned}
    \dot{x}_k &= f_k(x_1, \hdots, h_M; \mathbf{w}), \quad 1 \leq k \leq m\\
    \dot{h}_k &= f_k(x_1, \hdots, h_M; \mathbf{w}), \quad m < k \leq M
\end{aligned}
\end{equation}
with parameters $\mathbf{w} \in \mathbb{R}^l$ and a desired reduced model
\begin{equation}\label{eq:genred2}
    g_k(x_k^{d_k}, \{x_1^{(d)}\}_{d=0}^{d_1-1}, \{x_2^{(d)}\}_{d=0}^{d_2-1}, \hdots, \{x_m^{(d)}\}_{d=0}^{d_m-1}) = 0, \quad 1 \leq k \leq m.
\end{equation}
with polynomial expressions $g_k$.}
\Output{A description of the set of solutions $\{\mathbf{w}^*\} \subseteq \mathbb{R}^l$ for which~\eqref{eq:hfm3} satisfies the reduced model~\eqref{eq:genred2}.}
\BlankLine
\For{$k\leftarrow 1$ \KwTo $m$}{
Compute $x_k, \dot{x}_k, \ddot{x}_k, \hdots, x_k^{(d_k)}$ as functions of $x_1, \hdots, h_M$ and $\mathbf{w}$ through repeated differentiation of~\eqref{eq:hfm3}.
}
\BlankLine
\BlankLine
\For{$k\leftarrow 1$ \KwTo $m$}{
$\overline{g}_k \leftarrow g_k$ \tcp*[h]{copy expression k in reduced model}\\
\For{$j\leftarrow 1$ \KwTo $m$}{
\For{$d\leftarrow 1$ \KwTo $d_j$}{
Substitute $x_j^{d}(x_1, \hdots, h_M; \mathbf{w})$ into $\overline{g}_k$.
}
}
}
\BlankLine
\BlankLine
\emph{$\overline{g}_k$ are polynomials in $x_1, \hdots, h_M$ and $\mathbf{w}$}\\
\For{$k\leftarrow 1$ \KwTo $m$}{
Rearrange $\overline{g}_k$ into the form
\begin{equation}
    \overline{g}_k(x_1, \hdots, h_M; \mathbf{w}) = \sum_{\bm{\alpha} \in I_k} p_k^{\bm{\alpha}}(\mathbf{w})x_1^{\alpha_1}\hdots h_M^{\alpha_M}, \quad 1 \leq k \leq m
\end{equation}
}
\BlankLine
Use mathematica or homotopy continuation methods to solve the (likely overdetermined) polynomial system
\begin{equation}
    p_k^{\bm{\alpha}}(\mathbf{w}) = 0, \quad \forall \ \bm{\alpha} \in I_k, \ 1 \leq k \leq m.
\end{equation}
Save all solutions $\mathbf{w}^* \in \mathbb{R}^l$ in a set and return to the user.
\caption{Reduction Checker}\label{algo_rc}
\end{algorithm}\DecMargin{1em}

Although our algorithm below is written for dynamical systems of a polynomial form, we remind the reader that the optimal set of model parameters $\mathbf{w}^*$ can be found through general nonlinear optimization methods. The additional structure of polynomial dynamical systems allows us to leverage the power of polynomial system solvers to efficiently verify the reductions of our learned models.

\FloatBarrier

\bibliographystyle{apsrev4-2}
\bibliography{bibliography}